\documentclass[10pt,a4paper]{article}

\usepackage{mathtext}
\usepackage[active]{srcltx}
\usepackage[T2A]{fontenc}
\usepackage[utf8]{inputenc}
\usepackage[english]{babel}
\usepackage{amsfonts}
\usepackage{amsmath}
\usepackage{amsthm}
\usepackage{amssymb}
\usepackage{amscd}
\usepackage{graphicx}
\usepackage{wrapfig}
\usepackage{hyperref}
\usepackage{dsfont}
\def\be{\begin{eqnarray}}
\def\ee{\end{eqnarray}}

\newcommand{\eq}[2]{\begin{equation} \label{e:#1} #2 \end{equation}}
\newcommand{\eql}[2]{\begin{multline} \label{e:#1} #2 \end{multline}}
\newcommand{\al}[1]{\begin{align} #1 \end{align}}
\newcommand{\re}[1]{(\ref{e:#1})}
\newcommand{\rf}[1]{Fig.\ref{fig:#1}}

\newcommand{\lrb}{\left(}
\newcommand{\rrb}{\right)}
\newcommand{\ltb}{\left<}
\newcommand{\rtb}{\right>}
\newcommand{\lsb}{\left[}
\newcommand{\rsb}{\right]}
\newcommand{\lfb}{\left\{}
\newcommand{\rfb}{\right\}}

\newcommand{\dl}[1]{{\Delta_{#1}}}
\newcommand{\N}{{\mathcal{N}}}
\newcommand{\bg}[2]{{{\gamma}_{#1}^{#2}}}

\allowdisplaybreaks

\numberwithin{equation}{section}

\title{Check of AGT Relation for
Conformal Blocks on Sphere}
\author{V. Alba$^{\star}$ \ \ and\ \ And.
Morozov$^{\dagger}$}
\date{\vspace{-0.3cm}\footnotesize{$^{\star\dagger}${\tt ITEP},
\\
\emph{117218 Moscow, Russia}\\
$^{\star}${\tt	 Landau Institute for Theoretical Physics RAS},\\
\emph{119334 Moscow, Russia}\\
$^{\star}${\tt Department of General and Applied Physics,\\ Moscow Institute
of Physics and Technology},\\
\emph{141700 Dolgoprudny, Moscow Reg., Russia}\\
$^{\dagger}${\tt Physical Department, Moscow State University},\\
\emph{119991 Moscow, Russia}\\
$^{\star}${\tt Bogolyubov Institute for Theoretical Physics NASU},\\
\emph{03680 Kyiv, Ukraine}}}

\begin{document}

\maketitle
\vspace{-8cm}
\hfill ITEP/TH-74/09
\vspace{6,7cm}
\begin{abstract}
The AGT conjecture identifying conformal blocks with the Nekrasov functions
is investigated for the spherical conformal blocks with more than 4 external legs. The diagram technique which arises in conformal block calculation involves propagators and vertices. We evaluated vertices with two Virasoro algebra descendants and explicitly checked the AGT relation up to the third order of the expansion  for the $5-$point and $6-$point conformal blocks on sphere confirming all the predictions of {\tt arXiv:0906.3219} relevant in this situation. We propose that $U(1)-$factor can be extracted from the matrix elements of the free field vertex operators. We studied
the $n-$point case, and found out that our results confirm the AGT conjecture up to the third order expansions.
\end{abstract}

\footnotetext[7]{V.Alba e-mail: alba@itp.ac.ru, alba@itep.ru,\quad And.Morozov e-mail: andrey.morozov@itep.ru}
\vspace{-0.5cm}
\tableofcontents

\section{Introduction}

Recently Alday, Gaiotto and Tachikawa (AGT) proposed a conjecture \cite{agt}
which implies that some Liouville correlation functions are equal to
an integral of the squared absolute value of the Nekrasov full partition functions.
In particular, this conjecture implies that conformal blocks \cite{bpz,alz1,alz2,df}
are equal to the instanton Nekrasov partition functions \cite{nek,fp,ny1,ny2,shd}. These functions are defined only in patches of the moduli space of the SW vacua \cite{sw1,sw2}. Actually the AGT conjecture in its present
form involves only comb-type conformal blocks with a peculiar choice of points on
the Riemann sphere. There is a vast literature devoted to checks and various discussions of the AGT relation \cite{agt}. In particular, the conjecture was explicitly checked for the several first terms of the expansion of the $4-$point conformal block on sphere \cite{agt,3m},
of the $1-$point conformal block on torus \cite{agt,pog,am} and it was discovered in various particular cases \cite{rp1} - \cite{rp_l}. It was proven in special cases \cite{fl,rp9}. Following \cite{agt} we check this conjecture for the $5-$point and $6-$point conformal blocks, calculating the conformal blocks and Nekrasov partition functions order by order.

Evaluating the instanton partition functions in $\mathcal{N}=2$ SYM by direct integration over the
instanton moduli space was quite a hard problem and remained unsolved for a long time. Though the
ADHM construction explicitly determines the moduli space, there was a problem of regularization of integrals over this space. The problem was fixed using the $\Omega$-background method. However, after regularization the instanton partition functions become dependent on the regularization parameters $\epsilon_1$ and $\epsilon_2$. These partition functions are called Nekrasov partition functions.

 There is a useful diagram technique \cite{son,mors,4m} to calculate the conformal blocks. This technique contains two elements: the propagator $D_{\alpha}$ which is inverse of the Shapovalov matrix (i.e. the propagator is the inverse matrix of the scalar product between fields from the same level in the Verma module) and the vertex $\bg{\alpha\beta}{\delta}$ which is the coefficient in the operator product expansion of two fields into the third one. We study some properties of these vertices and calculate some of them.

The conformal block is represented as a series in projective invariants $x_i$ and is a function of conformal dimensions $\Delta=\alpha(Q-\alpha)$ and of the central charge $c=1+6Q^2$. On the other hand, the instanton partition function is a series in $q_i$ (exponents of complex coupling $\tau$) and is a function of Higgs' v.e.v. $a_i$, fundamental masses $\mu_i$, bifundamental masses $m_i$ and deformation parameters $\epsilon_{1},\epsilon_{2}$.
These two functions depend on different parameters. Hence, there should be
relations between these parameters.

We checked explicitly coincidence between
the conformal block and the instanton partition function in the case of the
5-point and the 6-point conformal blocks on sphere. These calculations allow
us to deduce the explicit formulae for the relations between $\alpha_i$ and
$\mu_i,m_i,\nu_i$. These formulae coincide with predictions of \cite{agt}.
Our consideration allows us to obtain the necessary relations
in the case of $n-$point conformal block on the sphere.
Due to the fact that the formulae for the conformal block and for the
Nekrasov partition function are factorized, the first three orders always reduce
to the $4-,5-,6-$point cases.

As soon as
we associate the Virasoro conformal block with the partition function for the $U(2)$-quiver
theories, the Nekrasov partition functions for the $SU(2)$-quiver theories
should be multiplied by a $U(1)-$factor. The explicit expression for this factor can be
extracted from the matrix elements of the free field vertex operators \cite{df,ff}.
For the free field case, the conformal block is equal to pure $U(1)-$factor.

\section{Nekrasov partition functions}
Neglecting higher derivatives one can write the low energy effective action
for $\mathcal{N}=2$ SYM with the aid of the so-called prepotential $\mathcal{F}$.
The prepotential consists of the three parts: classical, perturbative and non-perturbative
(instantonic part). The explicit answers for classic and perturbation part \cite{pert}  were
calculated a long time ago and the exact answer for the instanton part was proposed
by N. Seiberg and E. Witten(SW) \cite{sw1,sw2} with the help of duality
arguments, also see \cite{swpr}-\cite{mswb}. A.Losev, G.Moore, N.Nekrasov and S.Shatashvili presented a two-parameter deformation of the SW prepotential \cite{lns1,mns1,lns2,mns2} in the form of LMNS integrals \cite{Moore:1997dj}-\cite{Losev:1999tu}. These
integrals were calculated by Nekrasov \cite{nek} and are called Nekrasov partition functions. The SW prepotential is
a limit of the free energy which corresponds to the Nekrasov partition function
\eq{limit}
{
\mathcal{F}=\mathop{\lim\limits_{\epsilon_1\longrightarrow
0}}_{\epsilon_2\longrightarrow 0} \epsilon_1\epsilon_2\ln \mathcal{Z}_{full}.
}
The full partition function is factorized
\eq{fact}
{
\mathcal{Z}_{full}(\epsilon_1,\epsilon_2)=Z_{classic}Z_{pert}Z_{inst}(\epsilon_1
,\epsilon_2).
}

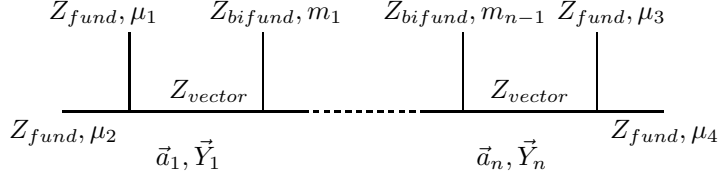
\begin{figure}[h]
%\begin{wrapfigure}{r}{275pt}
\begin{picture}(275,60)(-135,0)
\put(-15,20){\line(1,0){90}}
\put(120,20){\line(1,0){90}}
\multiput(75,20)(4,0){12}{\line(1,0){2}}
\multiput(10,20)(50,0){2}{\line(0,1){30}}
\multiput(135,20)(50,0){2}{\line(0,1){30}}
\put(-35,10){$Z_{fund},\mu_2$}
\put(25,25){$Z_{vector}$}
\put(145,25){$Z_{vector}$}
\put(20,0){$\vec{a}_1,\vec{Y}_1$}
\put(140,0){$\vec{a}_{n},\vec{Y}_{n}$}
\put(190,10){$Z_{fund},\mu_4$}
\put(-20,55){$Z_{fund},\mu_1$}
\put(105,55){$Z_{bifund},m_{n-1}$}
\put(40,55){$Z_{bifund},m_1$}
\put(170,55){$Z_{fund},\mu_3$}
\end{picture}
\caption{\label{fig:quiv}This diagram defines the instanton partition function for the
$\bigotimes\limits_{i=1}^{n}U(2)$ linear quiver theory. There is simple
correspondence between the quiver theories and the diagrams.}
%\end{wrapfigure}
\end{figure}
Nekrasov partition functions $\mathcal{Z}_{full}$ \cite{nek} was evaluated by F.Fucito, J.Morales, R.Poghossian \cite{fmp} for quiver theories corresponding to the orbifold projections of the $\N=4$ theories. The AGT conjecture implies relations between the comb-type conformal blocks and the partition function for the linear quiver theory. For a linear quiver with the $\bigotimes\limits_{i=1}^{n} U(2)$ gauge groups, the partition function (\rf{quiv}) is
\eql{nek1}
{
Z_\text{inst}= \sum_{\vec Y_1, \vec Y_2,\ldots, \vec Y_n}
\left(\prod_{i=1}^n q_i^{|\vec Y_i|} Z_\text{vector}(\vec a_i,\vec Y_i) \right)
Z_\text{fund}(\vec a_1,\vec Y_1,\mu_1)
Z_\text{fund}(\vec a_1,\vec Y_1,\mu_2)\times \\
\times \left(\prod_{i=1}^{n-1} Z_\text{bifund}(\vec a_i,\vec Y_i;\vec
a_{i+1},\vec Y_{i+1};m_i) \right)
Z_\text{fund}(\vec a_n,\vec Y_n,\mu_3)
Z_\text{fund}(\vec a_n,\vec Y_n,\mu_4),
}
where $\vec a_i=(a_{i,1},a_{i,2})$
is the diagonal of the adjoint scalar,
$\vec Y_i=(Y_{i1},Y_{i2})$ is the pair of the Ferrer-Young diagrams (\rf{fyd})
specifying the fixed instanton and
\eq{qpar}
{
q_i=e^{2\pi i \tau_i},\ \tau_i=\frac{4\pi i}{g^2_i}+\frac{\theta_i}{2\pi}.
}
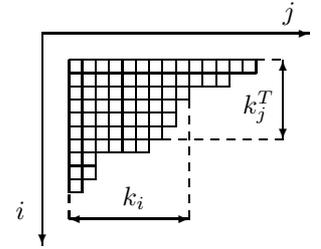
\begin{wrapfigure}{r}{150pt}
\begin{picture}(150,80)(-15,00)
\put(10,80){\vector(1,0){100}}
\put(10,80){\vector(0,-1){80}}
\put(100,85){$j$}
\put(00,10){$i$}
\put(20,20){\line(0,1){50}}
\put(25,20){\line(0,1){50}}
\put(30,25){\line(0,1){45}}
\put(35,35){\line(0,1){35}}
\put(40,35){\line(0,1){35}}
\put(45,35){\line(0,1){35}}
\put(50,35){\line(0,1){35}}
\put(55,40){\line(0,1){30}}
\put(60,45){\line(0,1){25}}
\put(65,55){\line(0,1){15}}
\put(70,60){\line(0,1){10}}
\put(75,60){\line(0,1){10}}
\put(80,60){\line(0,1){10}}
\put(85,65){\line(0,1){5}}
\put(90,65){\line(0,1){5}}
\put(20,70){\line(1,0){70}}
\put(20,65){\line(1,0){70}}
\put(20,60){\line(1,0){60}}
\put(20,55){\line(1,0){45}}
\put(20,50){\line(1,0){40}}
\put(20,45){\line(1,0){40}}
\put(20,40){\line(1,0){35}}
\put(20,35){\line(1,0){30}}
\put(20,30){\line(1,0){10}}
\put(20,25){\line(1,0){10}}
\put(20,20){\line(1,0){5}}
\multiput(65,10)(0,6){8}{\line(0,1){3}}
\multiput(20,10)(0,6){2}{\line(0,1){3}}
\multiput(55,40)(6,0){8}{\line(1,0){3}}
\multiput(90,70)(6,0){2}{\line(1,0){3}}
%\put(20,20){\dashbox{1}(75,30){}}
\put(100,40){\vector(0,1){30}}
\put(100,70){\vector(0,-1){30}}
\put(65,10){\vector(-1,0){45}}
\put(20,10){\vector(1,0){45}}
\put(85,50){$k^T_j$}
\put(40,15){$k_i$}
\end{picture}
\caption{\label{fig:fyd} Ferrer-Young diagram $[14,12,9,8,8,7,6,2,2,1]$, hhere
$s=(i,j)$ is a multiindex (coordinate on Ferrer-Young diagram) and
$k^T_j(Y),k_i(Y)$ are the height of column and length of row in Ferrer-Young
diagram correspondingly ($i=3,j=7$ in the picture). }
\end{wrapfigure}
$m_i$ is the mass of the bifundamental hypermultiplet charged under $SU(2)_i$
and $SU(2)_{i+1}$. $\mu_{1,2,3,4}$ are the masses of the fundamentals, $g_i$ is the coupling constant for $i-$th $U(2)$ group. Since the instanton partition function is factorized one can naturally associate it with the figure. One can write the partition function for a theory, if an external leg is connected with another external leg and with an internal one, then we associate this leg with $Z_{fund}(\vec{a},\vec{Y},\mu)$. If the external leg is connected with two internal legs then we associate this leg with
$Z_{bifund}(\vec{a}_i,\vec{Y}_i,\vec{a}_{i+1},\vec{Y}_{i+1},m_i)$, and internal
leg is associated with $Z_{vector}(\vec{a},\vec{Y})$. Each internal leg carries one $\vec{a}_i$ and one $\vec{Y}_i$. For instance, the linear quiver is associated with \rf{quiv}.
Now we define the bifundamental contribution
\eql{bifund}
{
Z_\text{bifund}(\vec a,\vec Y;\vec b,\vec W;m)
=\prod_{i,j=1}^2
\prod_{s\in Y_i}( E(a_i-b_j,Y_i,W_j,s)- m)\times
\\
\times\prod_{t\in W_j}(\epsilon -
E(b_j-a_i,W_j,Y_i,t)- m).
} Here
\eq{defE}
{
E(a,Y_1,Y_2,s)=a +\epsilon_1\lrb k^T_j(Y_1)-i+1\rrb -\epsilon_2(k_i(Y_2)-j),
}
where $k^T_j(Y),k_i(Y)$ is the height of column and length of row in the
Ferrer-Young diagram (\rf{fyd}) and $\epsilon=\epsilon_1+\epsilon_2$. Now it is easy to
define contributions of other hypermultiplets:
\begin{itemize}
\item adjoint hypermultiplet
\eq{adj}
{
Z_\text{adj}(\vec a,\vec Y,m)=Z_\text{bifund}(\vec a,\vec Y,\vec a,\vec Y,m),
}
\item vector multiplet
\eq{vec}
{
Z_\text{vector}(\vec a,\vec Y)=\frac{1}{Z_\text{adj}(\vec a,\vec Y,0)},
}
\item fundamental hypermultiplet
\eq{fundamental}
{
Z_{\text{fund}}(\vec a,\vec Y,m)=\prod_{i=1}^2 \prod_{s\in Y_i}(\phi(a_i,s) +
m),
}
\end{itemize}
where
\eq{phidef}
{
\phi(a,s)=a+\epsilon_1(i-1)+\epsilon_2(j-1).
}
There are exact rules how to write instanton part of the Nekrasov partition
function. With each external leg one associates fundamental hypermultiplet, with each integral channel one
associates the vector multiplet and with each leg which separating two internal channels one associates
the bifundamental hypermultiplet.

\section{CFT input}

Conformal field theory (CFT) \cite{pol1,pol2,bpz} studies local fields $A_i(z_i,\bar{z}_i)$ and
their correlators $\left<A_1(z_1,\bar{z}_1)A_2(z_2,\bar{z}_2)A_3(z_3,\bar{z}_3).
... \right>$ which depend on $z_i$, $\bar{z}_i$ and dimensions of these fields,
and they are linear functions with respect to each $A_i(z_i,\bar{z}_i)$.
The local fields form an operator algebra with product of the following form:
\eq{OPE}
{
A_i(z_i,\bar{z_i}) A_j (z_j,\bar{z_j}) =\sum \limits_{k}
C^k_{ij}(z_i,\bar{z_i},z_j,\bar{z_j}) A_k(z_j,\bar{z_j}).
}
From the translation and conformal invariances one gets
\eq{e1}
{
\begin{array}{l}
(\partial_{z_1}+\partial_{z_2})C^k_{12}(z_1,\bar{z}_1,z_2,\bar{z}_2)=0
\Rightarrow
C^k_{12}(z_1,\bar{z}_1,z_2,\bar{z}_2)=C^k_{12}(z_{12},\bar{z}_{12}) \\
(z_1\partial_{z_1}+z_2\partial_{z_2})C^k_{12}(z_{12},\bar{z}_{12})=(\Delta_k
-\Delta_1-\Delta_2)C^k_{12}(z_{12},\bar{z}_{12})\Rightarrow
C^k_{12}(z_{12},\bar{z}_{12})=\dfrac{C^k_{12}}{|z_{12}|^{
\Delta_1+\Delta_2-\Delta_k}},
\end{array}
}
where $z_{12}=z_1-z_2$ and $C_{12}^k$ are complex numbers.

Also in CFT one postulates the existence of a special operator, the stress-tensor $T(z)$
and conjugate operator $\bar{T}(\bar{z})$,
but we study only the stress-tensor itself and thus only the holomorphic part of
functions and fields. The stress-tensor operator defines the Virasoro operator algebra in the following
way. One defines the Virasoro operators using the Laurent series for the stress tensor \cite{bpz,df}:

\eq{Ldef}
{
T(u)A(z)=\sum\limits_{n=-\infty}^{\infty} \frac{L_n(z)}{(u-z)^{n+2}} A(z).
}

From this formula one also can get the integral definition of Virasoro operators:
\be
\left<L_{-n}A_1(0)A_2(z_2)A_3(z_3)... \right>=
\oint\limits_{0}\frac{dx}{x^{n-1}}\left<T(x)A_1(0)A_2(z_2)A_3(z_3)... \right>
\ee

Using OPE for the stress-tensor one can find out that these Virasoro operators satisfy with the commutation relation:
\be
\label{cr}
\left[ L_m,L_n\right]=(m-n) L_{m+n} +\frac{c}{12}(m^3-m)\delta_{m+n}.
\ee

The field is called primary if it is an eigenvector of the zero-th Virasoro operator
$L_0V=\Delta V$ and $L_nV=0,\ \forall n>0$, $\Delta$ is called the dimension
of this field.
One can find out using commutation relation (\ref{cr}) that
$L_0L_{-n}V=(\Delta+n)L_{-n}V$.
In CFT one studies the fields made from the primary fields using the Virasoro operators in the following way:
$V_{des}=L_{-k_N}.. L_{-k_2}L_{-k_1}V=L_{-Y}V$, they are called descendants.
$Y=\{k_1\geq k_2\geq... \geq k_N\}$ is called Ferrer-Young diagram
(\rf{fyd}).
The set of all descendants made from one primary field is called Verma module of
this field.

We will use the following parametrization \cite{df,ff} which is convenient for the formulation of the AGT conjecture:
\be
\label{e:param}
\Delta_{\alpha}=\frac{\alpha(\epsilon-\alpha)}{\epsilon_1\epsilon_2}
\\
c=1+\frac{6\epsilon^2}{\epsilon_1\epsilon_2}\label{e:pod}
\ee
where $\epsilon=\epsilon_1+\epsilon_2$.
%\newpage
\subsection{Diagram technique}
\begin{figure}[h]
\begin{picture}(220,50)(-100,-15)
\multiput(20,0)(150,0){2}{\line(1,0){95}}
\multiput(110,0)(4,0){20}{\line(1,0){2}}
\multiput(50,0)(40,0){2}{\line(0,1){30}}
\multiput(200,0)(40,0){2}{\line(0,1){30}}
\put(25,5){$1$}
\put(255,5){$1$}
\put(65,5){$D_{\beta_1}$}
\put(170,5){$D_{\beta_{n-2}}$}
\put(210,5){$D_{\beta_{n-1}}$}
\put(210,5){$D_{\beta_{n-1}}$}
\put(45,-15){$\bg{\alpha_0\beta_0}{\beta_1}$}
\put(85,-15){$\bg{\alpha_1\beta_1}{\beta_2}$}
\put(193,-15){$\bg{\alpha_{n-2}\beta_{n-2}}{\beta_{n-1}}$}
\put(237,-15){$\bg{\alpha_{n-1}\beta_{n-1}}{\beta_n}$}
\end{picture}
\caption{\label{fig:diagtech}Diagram technique. }
\end{figure}
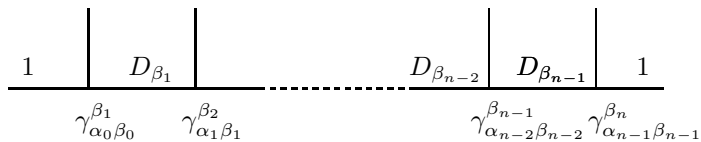
Diagram technique for the conformal block contains the two elements: the propagator $D_{\alpha}(Y,Y')$ and the triple vertices $\bg{\alpha_1\beta_1}{\beta_2}$ (\rf{diagtech}). Now we are going to describe the properties of these two elements.
\subsubsection{Propagator: the inverse Shapovalov matrix}

Another object in CFT is the scalar product of two fields
$\left<V_{\widehat{\alpha}}|V_{\widehat{\beta}}\right>$,
$\widehat{\alpha}=\{\alpha,Y_{\alpha}\}$. The scalar product is by definition an
object which is linear with respect to the fields and
obeys the following equation:
\be
\label{e:scprd}
\left<L_{-n}V_{\widehat{\alpha}}|V_{\widehat{\beta}}\right>=\left<V_{\widehat{
\alpha}}|L_{n}V_{\widehat{\beta}}\right>
\ee
As an example, one can define the scalar product as
\be\label{e:scalpr}
\left<V_{\hat\alpha} | V_{\hat\beta}\right>=
\left< V_{\hat\alpha}(0)V_{\hat\beta}(\infty)\right>.
\ee
since using the integral definition of Virasoro operators, one can get
\eql{intvir}
{
\left<L_{-n}V_{\Delta}(0)L_{-Y}V_{\Delta}(\infty) \right>=
\oint\limits_{0}\frac{dx}{x^{n-1}}\left<T(x)V_{\Delta}(0)L_{-Y'}V_{\Delta}
(\infty) \right>=\\
=\oint\limits_{\infty}\frac{dx x^{k-2}}{x^{n-1}}\left< V_{\Delta}(0)L_k
L_{-Y'}V(\infty)\right>=
\left< V_{\Delta}(0)L_{n}L_{-Y'}V_{\Delta}(\infty)\right>.
}
In our calculations we need the matrix of the Shapovalov form, defined as
\eq{sm}
{
H_{\hat\alpha \hat\beta}=\left<V_{\hat\alpha} | V_{\hat\beta}\right>
}
Using (\ref{e:scalpr}) one can find out that for two primaries the Shapovalov form
can be represented as
\eq{sp}
{
H_{\hat\alpha\hat\beta}=\left<L_{-Y_{\alpha}}V_{\alpha}|L_{-Y_{\beta}}V_{\beta}
\right>=
Q_{\Delta_{\alpha}}(Y_{\alpha},Y_{\beta}
)\delta_{\alpha\beta}
}
where $Q_{\Delta}(Y_{\alpha},Y_{\beta})$  has the block-diagonal form:
\eq{shfr3}
{
\begin{array}{l||c|c|cc|ccc|c}
Y_{\alpha} \backslash Y_{\beta} & \varnothing & [1] & [2] & [1,1] & [3] & [2,1]
& [1,1,1] &\ldots\\
\hline\hline
\varnothing &1&&&&&&&\\
\hline
[1] &&2\Delta&&&&&&\\
\hline
[2] &&&\frac{1}{2}(8\Delta + c)&6\Delta&&&&\\

[1,1] &&&6\Delta&4\Delta(1+2\Delta)&&&&\\
\hline
[3] &&&&&6\Delta+2c&2(8\Delta+c)&24\Delta&\\

[2,1] &&&&&2(8\Delta+c)&8\Delta^2+(34+c)\Delta + 2c&36\Delta(\Delta+1)&\\

[1,1,1] &&&&&24\Delta&36\Delta(\Delta+1)&24\Delta(\Delta+1)(2\Delta+1)&\\
\hline
\vdots&&&&&&&&\ddots\\
\end{array}.
}

Because of its block-diagonal form we can evaluate the inverse matrix of the Shapovalov form
$D_{\Delta}(Y_{\alpha},Y_{\beta})=Q_{\Delta}^{-1}(Y_{\alpha},Y_{\beta})$ which
we need for our purposes.
It also has block-diagonal form and each block is the inverse of the
corresponding block from $Q_{\Delta}$.

\subsubsection{Triple vertices}

There are two more objects within this framework. There are triple vertices
\al
{
\Gamma_{\phi\chi\psi}(Y_{\phi},Y_{\chi},Y_{\psi})&=
\left<L_{-Y_{\phi}}V_{\phi}(0)L_{-Y_{\chi}}V_{\chi}(1)L_{-Y_{\psi}}V_{\psi}
(\infty)\right>\\
\overline{\Gamma}_{\chi\phi}^{\psi}(Y_{\chi},Y_{\phi},Y_{\psi})&=
\left<L_{-Y_{\psi}}V_{\psi}|L_{-Y_{\chi}}V_{\chi}(1)L_{-Y_{\phi}}V_{\phi}
(0)\right>.
}
In principle these two triple vertices are different objects, but if we define
scalar product as in (\ref{e:scalpr})
they are equal to each other:
\eql{gmtre}
{
\overline{\Gamma}_{\chi\phi}^{\psi}(Y_{\chi},Y_{\phi},Y_{\psi})=
\left<L_{-Y_{\psi}}V_{\psi}|L_{-Y_{\chi}}V_{\chi}(1)L_{-Y_{\phi}}V_{\phi}
(0)\right>=\\
=\left<L_{-Y_{\phi}}V_{\phi}(0)L_{-Y_{\chi}}V_{\chi}(1)L_{-Y_{\psi}}V_{\psi}
(\infty)\right>=
\Gamma_{\phi\chi\psi}(Y_{\phi},Y_{\chi},Y_{\psi})
}

Using \re{intvir} and \re{scprd} one can find out that

\eql{gamrel}
{
\left<L_{-n}V_{\hat{\psi}}|V_{\hat{\chi}}(1)V_{\hat{\phi}}(0)\right>=
\sum\limits_{k>0}\frac{(n+1)!}{(k+1)!(n-k)!}\left<V_{\hat{\psi}}|(L_kV_{\hat{
\chi}})(1)V_{\hat{\phi}}(0)\right>
+\\
+(n+1)\left<V_{\widehat{\psi}}|(L_0V_{\widehat{\chi}})(1)V_{\widehat{\phi}}
(0)\right>+
\left<V_{\widehat{\psi}}|(L_{-1}V_{\widehat{\chi}})(1)V_{\widehat{\phi}}
(0)\right>+
\left<V_{\widehat{\psi}}|V_{\widehat{\chi}}(1)(L_nV_{\widehat{\phi}})(0)\right>
}

From this formula any $\overline{\Gamma}$ could be found in principle, even if some of them
are quite hard to evaluate in practice \cite{4m}.
There is a similar formula for $\Gamma$.
In fact, in conformal block one needs not $\Gamma$ and $\overline{\Gamma}$ but
$\gamma$ and $\overline{\gamma}$ defined as
\al
{
\Gamma_{\phi\chi\psi}(Y_{\phi},Y_{\chi},Y_{\psi})&=
\gamma_{\phi\chi\psi}(Y_{\phi},Y_{\chi},Y_{\psi})\left<V_{\phi}(0)V_{\chi}(1)V_{
\psi}(\infty)\right>
\\
\overline{\Gamma}_{\chi\phi}^{\psi}(Y_{\chi},Y_{\phi},Y_{\psi})&=
\overline{\gamma}_{\chi\phi}^{\psi}(Y_{\chi},Y_{\phi},Y_{\psi})\left<V_{\psi}|V_
{\chi}(1)V_{\phi}(0)\right>
}
For a special set of there are exact general for $\gamma$ and $\overline{\gamma}$
formulae:
\be
\gamma_{\phi\chi\psi}(Y_{\phi},\varnothing,\varnothing)=
\prod\limits_i\left(\Delta_{\phi}+k_i\Delta_{\chi}-\Delta_{\psi}+\sum\limits_{
j<i}k_j\right)
\\
\overline{\gamma}_{\chi\phi\psi}(\varnothing,\varnothing,Y_{\psi})=
\prod\limits_i\left(\Delta_{\psi}+k_i\Delta_{\chi}-\Delta_{\phi}+\sum\limits_{
j<i}k_j\right)
\ee
List of some other evaluated $\overline{\gamma}$ with the aid of \re{gamrel} is
given in the Appendix A.

We pay the much attention on these vertices because we use them to evaluate
conformal blocks.
More concretely, we use \re{OPE} to reduce the $n-$point conformal block to
the $(n-1)-$point conformal block. However, we do not know the structure constants
$C^k_{ij}$ in terms of dimensions. We can express them in terms of these
vertices
\eq{strcns}
{
\bar\Gamma_{\hat\psi\hat\phi}^{\hat\chi}\mathop{=}^{def} \ltb
V_{\hat\chi}\big|V_{\hat\psi}(1)V_{\hat\phi}(0)\rtb=\sum\limits_{\hat\xi}C_{
\hat\psi\hat\phi}^{\xi}H_{\hat\chi\hat\xi}
\Longrightarrow
C_{\hat\psi\hat\phi}^{\hat\xi}=\bar\Gamma_{\hat\psi\hat\phi}^{\hat\chi}\lrb
H^{-1}\rrb_{\hat\chi\hat\xi}.
}
One can rewrite this formula in the form
\eq{stcnst}
{
C_{\hat\psi\hat\phi}^{\hat\xi}=C_{\psi\phi}^{\xi}\bg{\psi\phi}{\chi}(Y_{\psi},Y_
{\phi},Y_{\chi})D_{\xi}(Y_{\chi},Y_{\xi})
}

\subsection{Conformal blocks}

First of all,  we do not really need to specify the conformal field theory because we will consider only conformal blocks which are completely fixed by the conformal invariance. For detailed definitions and properties, see \cite{bpz,cfzz}. The conformal block can be associated with some diagram. Given a set of fields at some points, one can construct different conformal blocks corresponding different
diagrams.

\subsubsection{The four-point conformal block}
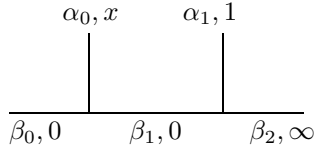
\begin{figure}[h]
%\begin{wrapfigure}{r}{130pt}
\begin{picture}(130,50)(-200,0)
\put(0,10){\line(1,0){110}}
\multiput(30,10)(50,0){2}{\line(0,1){30}}
\put(20,45){$\alpha_0,x$}
\put(65,45){$\alpha_1,1$}
\put(0,0){$\beta_0,0$}
\put(45,0){$\beta_1,0$}
\put(90,0){$\beta_2,\infty$}
\end{picture}
%\end{wrapfigure}
\end{figure}
One always can fix three points with the help of projective
symmetry. For $z_1=0,z_2=1,z_3=\infty$ the conformal block depends only on one projective
invariant $x$. One can use \re{OPE} and obtain
\eq{4c}
{
\left<V_{\alpha_0}(x)V_{\beta_0}(0)V_{\alpha_1}(1)V_{\beta_2}(\infty)
\right>=x^{-(\Delta_{\alpha_1}+\Delta_{\beta_0})}\sum\limits_{\hat\beta_1,
\hat\beta'}x^{\Delta_{\hat\beta_1}} \bar
\Gamma_{\alpha_0\beta_0}^{\hat\beta_1}(H^{-1})^{\hat\beta_1\hat\beta'}\Gamma_{
\hat\beta' \alpha_1\beta_2}.
}

With the help of \re{stcnst}, \re{OPE} one can rewrite \re{4c}
\eql{4cb}
{
\left<V_{\alpha_0}(x)V_{\beta_0}(0)V_{\alpha_1}(1)V_{\beta_2}(\infty) \right>=\\
=x^{-(\Delta_{\alpha_0}+\Delta_{\beta_0})}
\sum\limits_{\beta_1} x^{\Delta_{\beta_1}}\lrb
C_{\alpha_0\beta_0}^{\beta_1}C_{\alpha_1\beta_1}^{\beta_2}
\rrb\sum\limits_{l_1=0}^{\infty}x^{l_1}\mathop{\sum\limits_{|Y_1|=l_1}}_{Y_1'}
\bg{\alpha_0\beta_0}{\beta_1}(\varnothing,Y_1',\varnothing)
D_{\beta_1}(Y_1',Y_1)\bg{\alpha_1\beta_1}{\beta_2}(\varnothing,Y_1,
\varnothing)=\\
=x^{-(\Delta_{\alpha_0}+\Delta{\beta_0})}
\sum\limits_{\beta_1} x^{\Delta_{\beta_1}}\lrb
C_{\alpha_0\beta_0}^{\beta_1}C_{\alpha_1\beta_1}^{\beta_2}\rrb
\mathcal{B}_{\alpha_0\beta_0\alpha_1\beta_2}\lrb Y_1\big|x\rrb,
} where $\mathcal{B}_{\alpha_0\beta_0\alpha_1\beta_2}\lrb Y_1\big|x\rrb$
is the conformal block and $D_{\beta_1}$ is the inverse Shapovalov matrix. This formula
defines the four-point conformal block as a series in $x$.

\subsubsection{The five-point conformal block}
There are several conformal blocks.
block. We consider only two of them. The first one coincides with the conformal
block proposed in \cite{agt}.

We choose $z_1=y, z_2=xy,z_3=0,z_4=1,z_5=\infty$.
\begin{figure}[h]
\begin{picture}(220,50)(-170,0)
\put(20,0){\line(1,0){150}}
\multiput(50,0)(40,0){3}{\line(0,1){30}}
\put(25,-10){$\beta_0,0$}
\put(65,-10){$\beta_1,0$}
\put(105,-10){$\beta_2,0$}
\put(140,-10){$\beta_3,\infty$}
\put(35,35){$\alpha_0,xy$}
\put(80,35){$\alpha_1,y$}
\put(120,35){$\alpha_2,1$}
\end{picture}
\end{figure}
Then the expansion for the conformal block has the form
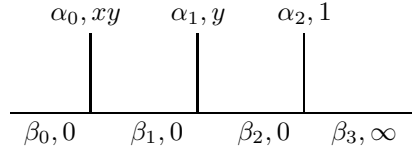
\eql{5cb}
{
\left<
V_{\alpha_1}(y)V_{\alpha_0}(xy)V_{\beta_0}(0)V_{\alpha_2}(1)V_{\beta_3}(\infty)
\right>=\\
=(xy)^{-(\alpha_0+\beta_0)}y^{-\alpha_1}\sum\limits_{\beta_1,\beta_2}x^{\Delta_{
\beta_1}}y^{\Delta_{\beta_2}}
C_{\alpha_0\beta_0}^{\beta_1}C_{\alpha_1\beta_1}^{\beta_2}C_{\alpha_2\beta_2}^{
\beta_3}\times\\
\times \sum_{l_1,l_2}x^{l_1}y^{l_2}
\mathop{\sum\limits_{|Y_1|=l_1, |Y_2|=l_2}}_{Y_1',Y_2'}
\bg{\alpha_0\beta_0}{\beta_1}(\varnothing,\varnothing,Y_{1}')D_{\beta_1}(Y_1',
Y_1)\bg{\alpha_1\beta_1}{\beta_2}(\varnothing,Y_1,Y_2')D_{\beta_2}(Y_2',Y_2)
\bg{\alpha_2\beta_2}{\beta_3}(\varnothing,Y_2,\varnothing).
}
This conformal block is a series in $x,y$. The conformal block is equal to the last
sum.

The another conformal block can be defined as
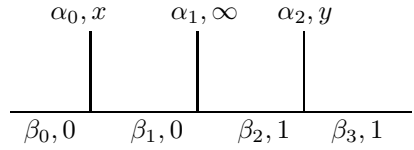
\begin{figure}[h]
\begin{picture}(220,50)(-170,0)
\put(20,0){\line(1,0){150}}
\multiput(50,0)(40,0){3}{\line(0,1){30}}
\put(25,-10){$\beta_0,0$}
\put(65,-10){$\beta_1,0$}
\put(105,-10){$\beta_2,1$}
\put(140,-10){$\beta_3,1$}
\put(35,35){$\alpha_0,x$}
\put(80,35){$\alpha_1,\infty$}
\put(120,35){$\alpha_2,y$}
\end{picture}
\end{figure}

\eql{5cbxy1}
{
\left<V_{\alpha_0}(x)V_{\beta_0}(0)V_{\beta_3}(1)V_{\alpha_2}(y)V_{\alpha_1}
(\infty)
\right>=x^{-(\Delta_{\alpha_0}+\Delta_{\beta_0})}(y-1)^{-(\Delta_{\alpha_3}
+\Delta_{\beta_{3}})}
\sum\limits_{\beta_1,\beta_2}x^{\Delta_{\beta_1}}(y-1)^{\Delta_{\beta_2}}
C_{\alpha_0\beta_1}^{\beta_2}C_{\alpha_2\beta_2}^{\beta_3}
C_{\beta_1\beta_2}^{\alpha_1}\times\\
\times
\sum\limits_{l_1,l_2}\mathop{\sum\limits_{|Y_1|=l_1,|Y_2|=l_2}}_{Y_1',Y_2'}x^{
l_1}(y-1)^{l_2} \bg{\alpha_0\beta_0}{\beta_1}(\varnothing,\varnothing,Y_1')
D_{\beta_1}(Y_1',Y_1)\bg{\beta_1\beta_2}{\alpha_1}(Y_1,Y_2',
\varnothing)\widetilde{D}_{\beta_2}(Y_2',Y_2)\bg{\alpha_2\beta_3}{\beta_2}
(\varnothing,Y_2,\varnothing)
}where $D_{\beta}(Y',Y)$ is the inverse Shapovalov form, while
$\widetilde{D}_{\beta}(Y,Y')$ is inverse deformed Shapovalov form, which means
that this is product of field not in 0 and $\infty$ but in $1$ and $\infty$.
\eq{dtil}
{
\widetilde{Q}_{\dl{}}([Y],[Y'])=\sum\limits_k\frac{1}{k!}Q_{\dl{}}\lrb\lsb
Y,1^k\rsb,[Y']\rrb.
}
One can find the explicit formulae for the first few levels in the Appendix.
The conformal block \re{5cbxy1} is a series in $x$ and $y-1$.
\subsubsection{The six-point conformal block}
We choose $z_1=z, z_2=yz,z_3=xyz,z_4=0,z_5=1, z_6=\infty$.
\begin{figure}[h]
\begin{picture}(220,50)(-150,0)
\put(20,0){\line(1,0){180}}
\multiput(50,0)(40,0){4}{\line(0,1){30}}
\put(15,-10){$\beta_0,0$}
\put(175,-10){$\beta_4,\infty$}
\put(60,-10){$\beta_1,0$}
\put(100,-10){$\beta_2,0$}
\put(140,-10){$\beta_3,0$}
\put(30,35){$\alpha_0,xyz$}
\put(75,35){$\alpha_1,yz$}
\put(120,35){$\alpha_2,z$}
\put(160,35){$\alpha_3,1$}
\end{picture}
\caption{\label{fig:6cb}This diagram defines six-point conformal block. }
\end{figure}
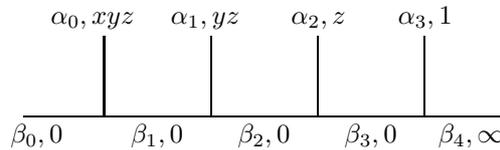
Then the conformal block (\rf{6cb}) has the form
\eql{6cb}
{
\ltb
V_{\alpha_0}(xyz)V_{\alpha_1}(yz)V_{\alpha_2}(z)V_{\beta_0}(0)V_{\alpha_3}(1)V_{
\beta_4}(\infty)\rtb=\\
=x^{-(\dl{\beta_0}+\dl{\alpha_0})}y^{-(\dl{\beta_0}+\dl{\alpha_0}+\dl{\alpha_1})
}z^{-(\dl{\beta_0}+\dl{\alpha_0}+\dl{\alpha_1}+\dl{\alpha_2})}
\sum\limits_{\beta_1,\beta_2,\beta_3}x^{\dl{\beta_1}}y^{\dl{\beta_2}}z^{\dl{
\beta_3}}C_{\alpha_0\beta_0}^{\beta_1}C_{\alpha_1\beta_1}^{\beta_2}C_{
\alpha_2\beta_2}^{\beta_3}C_{\alpha_3\beta_3}^{\beta_4}\times\\
\times
\sum_{l_1,l_2,l_3}x^{l_1}y^{l_2}z^{l_3}\mathop{\sum_{|Y_1|=l_1,|Y_2|=l_2,
|Y_3|=l_3}}_{Y_1',Y_2'Y_3'} \times\\
\times\bg{\alpha_0\beta_0}{\beta_1}(\varnothing,\varnothing,Y_1')D_{\beta_1}
(Y_1',Y_1)\bg{\alpha_1\beta_1}{\beta_2}(\varnothing,Y_1,Y_2')
D_{\beta_2}(Y_2',Y_3)\bg{\alpha_2\beta_2}{\beta_3}(\varnothing,Y_2,Y_3')
D_{\beta_3}(Y_3',Y_3)\bg{\alpha_3\beta_3}{\beta_4}(\varnothing,Y_3,\varnothing)
}
\subsubsection{The n-point conformal block}
Formula for the $n$-point conformal block (\rf{ncb})
\eq{cbnpre}
{
\mathcal{B}=\sum\limits_{l_1,\dots,l_n}x_1^{l_1}\dots
x_n^{l_n}\mathcal{B}^{(l_1,\dots,l_n)}
}
\eq{ncb}
{
\mathcal{B}^{(l_1,\dots,l_n)}
=\mathop{\mathop{\sum\limits_{|Y_1|=l_1}}_{\dots}}_{|Y_N|=l_n}\prod\limits_{i=1}
^{n+1}
\sum\limits_{Y_i'}
\bg{i-1\beta_{i-1}}{\beta_i}(\varnothing,Y_{i-1},Y_i')D_{\beta_i}(Y_i',Y_i),
}
where $x_{n+1}=1,l_{n+1}=0,Y_0\equiv\varnothing,Y_{n+1}\equiv\varnothing$,
$D_{\beta_i}$ is the inverse Shapovalov form.

\subsection{Explicit patch of the moduli space}
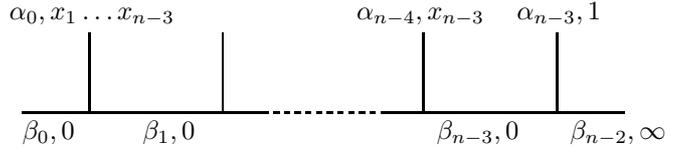
\begin{wrapfigure}{r}{250pt}
\begin{picture}(250,65)(-20,0)
\label{f:fd}
\put(-15,20){\line(1,0){90}}
\put(120,20){\line(1,0){90}}
\multiput(75,20)(4,0){12}{\line(1,0){2}}
\multiput(10,20)(50,0){2}{\line(0,1){30}}
\multiput(135,20)(50,0){2}{\line(0,1){30}}
\put(-15,10){$\beta_0,0$}
\put(30,10){$\beta_1,0$}
\put(140,10){$\beta_{n-3},0$}
\put(190,10){$\beta_{n-2},\infty$}
\put(-20,55){$\alpha_0,x_1\dots x_{n-3}$}
\put(110,55){$\alpha_{n-4},x_{n-3}$}
\put(170,55){$\alpha_{n-3},1$}
\end{picture}
\caption{\label{fig:ncb}This diagram defines the fusing rule for fields in
the conformal block. We called this diagram type comb-type diagram. AGT
conjecture speculates only about diagram of this type. }
\end{wrapfigure}

Because of the $SL(2,\mathds{Z})$ symmetry n-point conformal block depends only
on $n-3$ projective invariants. The conformal block is well defined for any non-coinciding coordinates.
However, the Nekrasov partition function is defined only in neighborhood of the infinity in the moduli space of SW vacua \cite{sw1,sw2}.

In this
patch the parameters $q_i$ must be small ($q_i\ll 1$). The Nekrasov partition
function is a series in $q_i$, while the conformal block is a series in
$x_i$-projective invariants. One should set $x_i=q_i$ modulo permutations.

 Consequently for the conformal block one should choose the fields in the special
points
\eq{sph}
{
\infty,1,x_{n-3},x_{n-3}x_{n-4},\dots, \prod\limits_{i=1}^{n-3} x_i,0,\quad
x_i\ll 1,
} where $n$ is the number of fields in the conformal block.
The choice turns out to be crucial. The CFT diagram defines the order of fusion, hence this special choice of the
projective invariants gives only one type of diagram that is the comb diagrams. In \cite{agt} it was
conjectured  about the comb type of the diagram with special hierarchy of points. For instance, the 5-point conformal block should be calculated only
for points $0,xy,y,1,\infty$, where $x,y\ll 1$. Other choices as we checked do
not provide the desired equality between the conformal block and the Nekrasov partition
function. In more detail, when one fuses the fields at points $y$ and 1, and the fields at points $x$ with the fields at points $0$, \rf{oddfig} $a)$ one does not
obtain the equality of the conformal block and the instanton partition
function. For the $6-$point conformal block, there is an extra diagram type, \rf{oddfig} $b)$.
\begin{figure}[h]
\begin{picture}(550,65)(-30,0)
\put(18,20){\line(1,0){80}}
\put(-1.5,40){\line(1,-1){20}}
\put(-1.5,0){\line(1,1){20}}
\put(97.5,20){\line(1,-1){20}}
\put(97.5,20){\line(1,1){20}}
\put(57,20){\line(0,1){30}}
\put(-17,45){$\alpha_1,x$}
\put(-17,-10){$\alpha_2,0$}
\put(107,45){$\alpha_3,y$}
\put(107,-10){$\alpha_4,1$}
\put(45,55){$\alpha_5,\infty$}
\put(25,25){$\beta_{1},0$}
\put(66,25){$\beta_{2},1$}
\put(300,15){\line(1,0){80}}
\put(300,15){\line(-1,1){20}}
\put(280,-5){\line(1,1){20}}
\put(380,15){\line(1,1){20}}
\put(400,-5){\line(-1,1){20}}
\put(340,15){\line(0,1){15}}
\put(340,30){\line(1,1){15}}
\put(340,30){\line(-1,1){15}}
\put(50,0){$a)$}
\put(340,0){$b)$}
\end{picture}
\caption{\label{fig:oddfig} a)This is the wrong fusion order. b) This type of the conformal block corresponds to "the generalized quiver". The correspondence in this case has not been discovered explicitly yet. }
\end{figure}
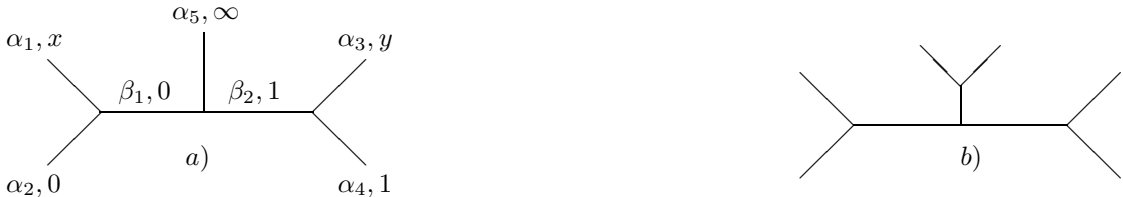

\section{$U(1)$-factor from the free fields conformal block}
Manifest expressions for the Nekrasov functions which we discuss in sec.2 are for the theory with
gauge group $\bigotimes\limits_{i=1}^{n} SU(N)$ (quiver theories), in our case
$\bigotimes\limits_{i=1}^{n} SU(2)$. At the same time, in the AGT conjecture the theories with gauge group $\bigotimes\limits_{i=1}^{n} U(2)$, hence one has
to multiply those partition functions by contributions of $U(1)$-factors.
In the original paper \cite{agt} there is the explicit formula for the $U(1)-$factor. We propose a general method to
reproduce these factors for any quiver theory. Let us consider the free field
correlator
\eq{ffc}
{
\ltb V_{\alpha_1}(z_1,\bar z_1)\dots V_{\alpha_n}(z_n,\bar z_n)\rtb=
\prod\limits_{i<j}|z_i-z_j|^{-2\alpha_i\alpha_j},
} where $V_{\alpha_i}(z_i,\bar z_i)=\colon e^{\alpha_i \varphi(z_i,\bar
z_i)}\colon$. We are going to consider only holomorphic part of this correlator (the conformal block with a restricted intermediate dimension). Assuming that monomial factors contribute only to the classical part,
we are only interested in non-monomial multipliers. This allows us to write the
recurrence relation
\eq{rr}
{
Z_{n}^{U(1)}=Z_{n-1}^{U(1)}\prod\limits_{i=1}^{n-3}\lrb
1-\prod\limits_{j=i}^{n-3}x_j\rrb^{-\nu_{i+n(n-1)/2}},
}
where $x_i$ are the projective invariants, which mean that with the aid of the conformal symmetry we can move three points to the points $0,1,\infty$, then other coordinates $z_i$ become $x_i$. $\nu_i$ is the power in $U(1)$-factor.
This recurrent relation can be solved by direct calculation:
\eq{u1}
{
\boxed{
Z_n^{U(1)}=\prod\limits_{i=1}^{n}\prod\limits_{j=1}^{i-3}\lrb 1-\prod\limits_{k=i-j+1}^{i-3}x_k\rrb^{-\nu_{j+i(i-1)/2}}
}
}
 We remind that $Z_{n-3}$ corresponds to the $n$-point conformal block. For
the four-point conformal block, the $U(1)-$factor is of the form $(1-x)^{-\nu}$.

\section{Relations between 2d conformal blocks and 4d Nekrasov partition
functions}
In this section we discuss the relations between the conformal
dimensions and the Nekrasov partition function parameters. We check the relation
\eq{atrel}
{
\mathcal{B}=\mathcal{Z}^{U(1)}\mathcal{Z}_{inst}
}where the conformal block is a function of the parameters
\eq{cbpar}
{
\mathcal{B}=\mathcal{B}(\Delta_1,\dots,\Delta_n,\Delta_{\beta_1},\dots,\Delta_{
\beta_{n-3}},x_1,\dots,x_{n-3}),
} while the Nekrasov partition function and $U(1)-$factor are the functions of

\eq{nekpar}
{
\begin{array}{l}
 \mathcal{Z}_{inst}=\mathcal{Z}_{inst}(\epsilon_1,\epsilon_2,\mu_1,\mu_2,\mu_3,
\mu_4,m_1,\dots,m_{n-3},a_1,\dots,a_{n-3},q_1,\dots,q_{n-3})\\
\mathcal{Z}^{U(1)}=\mathcal{Z}^{U(1)}(\nu_1,\dots,\nu_{n(n+3)/2},q_1,\dots,q_{
n-3}. )
\end{array}.
}
We use the special parametrization for the conformal dimensions and central charge
\re{pod}, because there is a linear relations between
$\alpha_i$ and $m_i,\mu_i,\epsilon_i$.
The intermediate dimensions are given by
\eq{idi}
{
\beta_i=a_i+\frac{\epsilon}{2}, i=1,\dots,n-3,
}
where $\epsilon=\epsilon_1+\epsilon_2$, and $\beta_i$ parametrize the intermediate
dimensions in accordance with \re{param}.
After substituting \re{idi} at each level of the conformal block expansion,its coefficients of the expansion
depend on $a_i$.

\subsection{Symmetries \label{symmsec}}

From \re{param} it is easy to see that there are symmetries in our answers with respect to
the reflection $\alpha_i\rightarrow(\epsilon-\alpha_i)$ of any external dimensions.
For all $n$-point conformal blocks except for the $4$-point one the variety of solutions is due to these symmetries.
Thus there are $2^n$ cases of the CFT side of solutions.
From \re{nek1} one can see also that the Nekrasov partition function would not change if one permutes
$\mu_1$ and $\mu_2$ or $\mu_3$ and $\mu_4$. So there are $4$ cases of the Nekrasov partition function side of solutions.
Thus for AGT conjecture in the $n$-point case there are $2^{n-2}$ solutions for parameters.

There is also exceptional case of the $4$-point conformal block.
In this case one has an additional symmetry $\alpha_{0}\leftrightarrow\beta_{0}$
and $\alpha_{1}\leftrightarrow\beta_{2}$ which doubles the number of solutions.
Thus there are $8$ different solutions for the $4$-point case.
The explanation of this symmetry is following:
the conformal block depends on variables, which are double ratios
$x=\dfrac{z_{\alpha_0\beta_0}z_{\beta_2\alpha_1}}{z_{\alpha_1\beta_0}z_{\beta_2\alpha_0}}$.
Thus for the $4$-point conformal block there is only one variable $x$ and it  would not change
if one applies this symmetry.
Thus the answer derived from the correct one using this permutation will also satisfy the AGT conjecture.
However for the general $n$-point conformal block there are several variables
and such symmetry can reserve only one of them.
Thus this symmetry does not exist for general $n$-point conformal block case.

\subsection{The four-point case}
In this case, the $U(1)-$factor is of the form $(1-x)^{-\nu}$.

Explicit expressions for the conformal block and for the Nekrasov partition
function expansion are
%\begin{wrapfigure}{r}{100pt}
$\begin{picture}(100,40)(-20,0)
\put(-10,20){\line(1,0){80}}
\multiput(15,20)(30,0){2}{\line(0,1){20}}
\put(-15,10){$\beta_{0},0$}
\put(20,10){$\beta_1,0$}
\put(50,10){$\beta_{2},\infty$}
\put(0,45){$\alpha_{0},x$}
\put(40,45){$\alpha_{1},1$}
\end{picture}$
%\end{wrapfigure}
\begin{itemize}
\item Level {\bf[1]}
\begin{itemize}
 \item Nekrasov partition function
\eq{4nek1}
{
\mathcal{Z}^{(1)}=Z\lrb[1],\varnothing\rrb+Z\lrb\varnothing,[1]\rrb+\nu=
-\frac{1}{\epsilon_1\epsilon_2}\cdot
\frac{\prod_{r=1}^4 (a + \mu_r)}
{2a(2a+\epsilon)} -\frac{1}{\epsilon_1\epsilon_2}\cdot
\frac{\prod_{r=1}^4 (a - \mu_r)}
{2a(2a-\epsilon)} +\nu.
}
 \item Conformal block

\eql{4cb1}
{
\mathcal{B}^{(1)}=
\bg{\alpha_0\beta_0}{\beta_1}(\varnothing,\varnothing,[1])D_{\beta_1}([1],[1]
)\bg{\alpha_1\beta_1}{\beta_2}(\varnothing,[1],\varnothing)=
\frac{(\dl{\beta_1}+\dl{\alpha_0}-\dl{\beta_0})(\dl{\beta_1}+\dl{\alpha_1}-\dl{
\beta_2})}{2\dl{\beta_1}}=\\
=\frac{2\lrb \dfrac{\epsilon^2}{4}-a^2
+\alpha_0(\epsilon-\alpha_0)-\beta_0(\epsilon-\beta_0)\rrb
\lrb\dfrac{\epsilon^2}{4}-a^2
+\alpha_1(\epsilon-\alpha_1)-\beta_2(\epsilon-\beta_2)\rrb}{
\epsilon_1\epsilon_2(\epsilon^2-4a^2)}
..
}
\end{itemize}
\item Level {\bf[2]}
\begin{itemize}
\item Nekrasov partition function
\eql{4nek2}
{
\mathcal{Z}^{(2)}=Z\lrb[2],\varnothing\rrb+Z\lrb\varnothing,[2]\rrb+Z\lrb[1,1],
\varnothing\rrb+Z\lrb\varnothing,[1,1]\rrb+\\
+Z\lrb[1],[1]\rrb+\nu\lrb
Z\lrb[1],\varnothing\rrb+Z\lrb\varnothing,[1]\rrb\rrb+\frac{\nu(\nu+1)}{2}.
}
\item Conformal block
\eql{4cb2}
{
\mathcal{B}^{(2)}=\bg{\alpha_0\beta_0}{\beta_1}(\varnothing,\varnothing,[2])D_{
\beta_1}([2],[2])\bg{\alpha_1\beta_1}{\beta_2}(\varnothing,[2],\varnothing)+\bg{
\alpha_0\beta_0}{\beta_1}(\varnothing,\varnothing,[2])D_{\beta_1}([2],[1^2])\bg{
\alpha_1\beta_1}{\beta_2}(\varnothing,[1^2],\varnothing)+
\\
+\bg{\alpha_0\beta_0}{\beta_1}(\varnothing,\varnothing,[1^2])D_{\beta_1}([1^2],[
2])\bg{\alpha_1\beta_1}{\beta_2}(\varnothing,[2],\varnothing)+\bg{
\alpha_0\beta_0}{\beta_1}(\varnothing,\varnothing,[1^2])D_{\beta_1}([1^2],[1^2]
)\bg{\alpha_1\beta_1}{\beta_2}(\varnothing,[1^2],\varnothing).
}
\end{itemize}
\item Level {\bf[3]}
\begin{itemize}
\item Nekrasov Partition function
\eql{4nek3}
{
\mathcal{Z}^{(3)}=Z\lrb[3],\varnothing\rrb+Z\lrb\varnothing,[3]\rrb+Z\lrb[2,1],
\varnothing\rrb+Z\lrb[1,1,1],\varnothing\rrb+
\\
+Z\lrb\varnothing,[2,1]\rrb+Z\lrb\varnothing,[1,1,1]\rrb+Z\lrb[2],[1]\rrb+Z\lrb[
1],[2]\rrb+Z\lrb[1,1],[1]\rrb+Z\lrb[1],[1,1]\rrb+
\\
+\nu\lrb
Z\lrb[2],\varnothing\rrb+Z\lrb\varnothing,[2]\rrb+Z\lrb[1,1],
\varnothing\rrb+Z\lrb\varnothing,[1,1]\rrb+Z\lrb[1],[1]\rrb\rrb
+\\
+\frac{\nu(\nu+1)}{2}\lrb
Z\lrb[1],\varnothing\rrb+Z\lrb\varnothing,[1]\rrb\rrb+\frac{\nu(\nu+1)(\nu+2)}{6
}.
}
\item Conformal block
\eql{4cb3}
{
\mathcal{B}^{(3)}=\bg{\alpha_0\beta_0}{\beta_1}(\varnothing,\varnothing,[3])D_{
\beta_1}([3],[3])\bg{\alpha_1\beta_1}{\beta_2}(\varnothing,[3],\varnothing)+\bg{
\alpha_0\beta_0}{\beta_1}(\varnothing,\varnothing,[3])D_{\beta_1}([3],[2,1])\bg{
\alpha_1\beta_1}{\beta_2}(\varnothing,[2,1],\varnothing)+
\\
+\bg{\alpha_0\beta_0}{\beta_1}(\varnothing,\varnothing,[3])D_{\beta_1}([3],[1^3]
)\bg{\alpha_1\beta_1}{\beta_2}(\varnothing,[1^3],\varnothing)+\bg{
\alpha_0\beta_0}{\beta_1}(\varnothing,\varnothing,[2,1])D_{\beta_1}([2,1],[3]
)\bg{\alpha_1\beta_1}{\beta_2}(\varnothing,[3],\varnothing)+
\\
+\bg{\alpha_0\beta_0}{\beta_1}(\varnothing,\varnothing,[2,1])D_{\beta_1}([2,1],[
2,1])\bg{\alpha_1\beta_1}{\beta_2}(\varnothing,[2,1],\varnothing)+\bg{
\alpha_0\beta_0}{\beta_1}(\varnothing,\varnothing,[2,1])D_{\beta_1}([2,1],[1^3]
)\bg{\alpha_1\beta_1}{\beta_2}(\varnothing,[1^3],\varnothing)+
\\
+\bg{\alpha_0\beta_0}{\beta_1}(\varnothing,\varnothing,[1^3])D_{\beta_1}([1^3],[
3])\bg{\alpha_1\beta_1}{\beta_2}(\varnothing,[3],\varnothing)+\bg{
\alpha_0\beta_0}{\beta_1}(\varnothing,\varnothing,[1^3])D_{\beta_1}([1^3],[2,1]
)\bg{\alpha_1\beta_1}{\beta_2}(\varnothing,[2,1],\varnothing)+
\\
+\bg{\alpha_0\beta_0}{\beta_1}(\varnothing,\varnothing,[1^3])D_{\beta_1}([1^3],[
1^3])\bg{\alpha_1\beta_1}{\beta_2}(\varnothing,[1^3],\varnothing).
}
\end{itemize}
\item etc.
\end{itemize}
More explicit formulae could be found in the Appendix.
In order to solve the system
\eq{4sys}
{
\left\{\begin{array}{l}
\mathcal{B}^{(1)}=\mathcal{Z}^{(2)}\\
\mathcal{B}^{(2)}=\mathcal{Z}^{(2)}
\end{array}\right.,
}
it is convenient to use other variables $s_i$
\al
{
\sigma_1&=\mu_1+\mu_2+\mu_3+\mu_4,
&\sigma_3&=\mu_1\mu_2\mu_3+\mu_1\mu_2\mu_4+\mu_1\mu_3\mu_4+\mu_2\mu_3\mu_4,
\notag\\
\sigma_2&=\mu_1\mu_2+\mu_1\mu_3+\mu_1\mu_4+
\mu_2\mu_3+\mu_2\mu_4+\mu_3\mu_4, &\sigma_4&=\mu_1\mu_2\mu_3\mu_4. \\
}

Then, the simple relations between the masses in 4d theory and the $\alpha$ parameters in 2d
CFT on sphere are
\eq{1}
{
\begin{array}{ll}
 \mu_1=-\frac{\epsilon}{2} +\alpha_0+\beta_0&\mu_2=\frac{\epsilon}{2}
+\alpha_0-\beta_0\\
 \mu_3=\frac{3\epsilon}{2} -\alpha_1-\beta_2&\mu_4=\frac{\epsilon}{2}
-\alpha_1+\beta_2
\end{array}, \nu=\frac{2\alpha_0(\epsilon-\alpha_1)}{\epsilon_1\epsilon_2},
}

Also using symmetries described in Sec.\ref{symmsec} one can obtain other solutions.
For this simple case one can count directly number of solutions,
and find out that there are exactly 8 different solutions in this case.

These solutions satisfy all other equations in the third order.

\subsection{The five-point case}
The $U(1)-$factor is of the form $(1-x)^{-\nu_1}(1-y)^{-\nu_2}(1-xy)^{-\nu_3}$.
Explicit expressions for the conformal block expansion and for the Nekrasov partition
function expansion:
\begin{itemize}
 \item Level {\bf [1,0]}
\begin{itemize}
\item Nekrasov partition function
\eql{5nek10}
{
\mathcal{Z}^{(1,0)}=
Z([[1],\varnothing],[\varnothing,\varnothing])+Z([\varnothing,[1]],[\varnothing,
\varnothing])+\nu_1=\\
=-\frac{(a_1+\mu_1)(a_1+\mu_2)(a_1-a_2+\epsilon-m_1)(a_1+a_2+\epsilon-m_1)}{
(2\epsilon_1\epsilon_2a_1(2a_1+\epsilon))}-
\\
-\frac{(-a_1+\mu_1)(-a_1+\mu_2)(-(a_1+a_2)+\epsilon-m_1)(-a_1+a_2+\epsilon-m_1)}
{(2\epsilon_1\epsilon_2a_1(2a_1-\epsilon))}+\nu_1
}
\item Conformal block
\eql{5cb10}
{
\mathcal{B}^{(1,0)}=\bg{\alpha_0\beta_0}{\beta_1}(\varnothing,\varnothing,[1])D_
{\beta_1}([1],[1]) \bg{\alpha_1\beta_1}{\beta_2}(\varnothing,[1],\varnothing)=
\frac{(\dl{\beta_1}+\dl{\alpha_0}-\dl{\beta_0})(\dl{\beta_1}+\dl{\alpha_1}-\dl{
\beta_2})}{2\dl{\beta_1}}=\\
=\frac{2\lrb
\dfrac{\epsilon^2}{4}
-a_1^2+\alpha_0(\epsilon-\alpha_0)-\beta_0(\epsilon-\beta_0)\rrb \lrb-a_1^2
+\alpha_1(\epsilon-\alpha_1)+a_2^2\rrb}
{\epsilon_1\epsilon_2(\epsilon^2-4a_1^2)}
..
}
\end{itemize}
\item Level {\bf [0,1]}
\begin{itemize}
\item Nekrasov partition function
\eql{5nek01}
{
Z([\varnothing,\varnothing],[[1],\varnothing])+Z([\varnothing,\varnothing],[
\varnothing,[1]])+\nu_2=\\
=-\frac{(a_2+\mu_3)(a_2+\mu_4)(-a_1+a_2+m_1)(a_1+a_2+m_1)}{
2\epsilon_1\epsilon_2a_2(2a_2+\epsilon)}-\\
-\frac{(-a_2+\mu_3)(-a_2+\mu_4)(a_1+a_2-m_1)(-a_1+a_2-m_1)}{
2\epsilon_1\epsilon_2a_2(2a_2-\epsilon)} +\nu_2
}
\item Conformal block
\eql{5cb01}
{
\mathcal{B}^{(0,1)}=\bg{\alpha_1\beta_1}{\beta_2}(\varnothing,\varnothing,[1])D_
{\beta_2}([1],[1]) \bg{\alpha_2\beta_2}{\beta_3}(\varnothing,[1],\varnothing)=
\frac{(\dl{\beta_2}+\dl{\alpha_1}-\dl{\beta_1})(\dl{\beta_2}+\dl{\alpha_2}-\dl{
\beta_3})}{2\dl{\beta_2}}=\\
=\frac{2\lrb -a_2^2 +\alpha_1(\epsilon-\alpha_1)+a_1^2\rrb
\lrb\dfrac{\epsilon^2}{4}-a_2^2
+\alpha_2(\epsilon-\alpha_2)-\beta_3(\epsilon-\beta_3)\rrb}{
\epsilon_1\epsilon_2(\epsilon^2-4a_2^2)}
}
\end{itemize}

\item Level {\bf [2,0]}
\begin{itemize}
\item Nekrasov partition function
\eql{5nek20}
{
Z\lrb \lsb [2],\varnothing\rsb,\lsb \varnothing,\varnothing\rsb \rrb+Z\lrb\lsb
[1],[1]\rsb,\lsb \varnothing,\varnothing\rsb\rrb+Z\lrb\lsb \varnothing,[2]\rsb
,\lsb \varnothing,\varnothing\rsb \rrb+\\+
Z\lrb\lsb [1,1],\varnothing\rsb,\lsb \varnothing,\varnothing\rsb \rrb+Z\lrb\lsb
\varnothing,[1,1]\rsb,\lsb \varnothing,\varnothing\rsb \rrb+\\
+\nu_1\lrb Z\lrb\lsb [1],\varnothing\rsb,\lsb \varnothing,\varnothing\rsb
\rrb+Z\lrb\lsb \varnothing,[1]\rsb,\lsb \varnothing,\varnothing\rsb
\rrb\rrb+\frac{1}{2}\nu_1(\nu_1+1).
}
\item Conformal block
\eql{5cb20}
{
\mathcal{B}^{(2,0)}=\bg{\alpha_0\beta_0}{\beta_1}(\varnothing,\varnothing,[2])D_
{\beta_1}([2],[2]) \bg{\alpha_1\beta_1}{\beta_2}(\varnothing,[2],\varnothing)+
\\
+\bg{\alpha_0\beta_0}{\beta_1}(\varnothing,\varnothing,[2])D_{\beta_1}([2],[1^2]
) \bg{\alpha_1\beta_1}{\beta_2}(\varnothing,[1^2],\varnothing)+
\\
+\bg{\alpha_0\beta_0}{\beta_1}(\varnothing,\varnothing,[1^2])D_{\beta_1}([1^2],[
2]) \bg{\alpha_1\beta_1}{\beta_2}(\varnothing,[2],\varnothing)+
\\
+\bg{\alpha_0\beta_0}{\beta_1}(\varnothing,\varnothing,[1^2])D_{\beta_1}([1^2],[
1^2]) \bg{\alpha_1\beta_1}{\beta_2}(\varnothing,[1^2],\varnothing).
}
\end{itemize}
\item Level {\bf [0,2]}
\begin{itemize}
\item Nekrasov partition function
\eql{5nek02}
{
Z\lrb\lsb \varnothing,\varnothing\rsb,\lsb [2],\varnothing\rsb\rrb+Z\lrb\lsb
\varnothing,\varnothing\rsb,\lsb \varnothing,[2]\rsb\rrb+Z\lrb\lsb
\varnothing,\varnothing\rsb,\lsb [1],[1]\rsb \rrb+\\
+Z\lrb\lsb \varnothing,\varnothing\rsb,\lsb [1,1],\varnothing\rsb\rrb+Z\lrb\lsb
\varnothing,\varnothing\rsb,\lsb \varnothing,[1,1]\rsb \rrb+\\
+\nu_2\lrb Z\lrb\lsb \varnothing,\varnothing\rsb,\lsb [1],\varnothing\rsb
\rrb+Z\lrb\lsb\varnothing,\varnothing\rsb
,\lsb\varnothing,[1]\rsb\rrb\rrb+\frac{1}{2}\nu_2(\nu_2+1)
}
\item Conformal block
\eql{5cb02}
{
\mathcal{B}^{(0,2)}=\bg{\alpha_1\beta_1}{\beta_2}(\varnothing,\varnothing,[2])D_
{\beta_2}([2],[2]) \bg{\alpha_2\beta_2}{\beta_3}(\varnothing,[2],\varnothing)+
\\
+\bg{\alpha_1\beta_1}{\beta_2}(\varnothing,\varnothing,[2])D_{\beta_2}([2],[1^2]
) \bg{\alpha_2\beta_2}{\beta_3}(\varnothing,[1^2],\varnothing)+
\\
+\bg{\alpha_1\beta_1}{\beta_2}(\varnothing,\varnothing,[1^2])D_{\beta_2}([1^2],[
2]) \bg{\alpha_2\beta_2}{\beta_3}(\varnothing,[2],\varnothing)+\\
+\bg{\alpha_1\beta_1}{\beta_2}(\varnothing,\varnothing,[1^2])D_{\beta_2}([1^2],[
1^2]) \bg{\alpha_2\beta_2}{\beta_3}(\varnothing,[1^2],\varnothing)
}
\end{itemize}
\item Level {\bf [1,1]}
\begin{itemize}
\item Nekrasov partition function
\eql{5nek11}
{
Z\lrb \big [ [1],\varnothing\big ],\big [ [1],\varnothing\big ]\rrb +Z\lrb
\big[\varnothing,[1]\big],\big [ [1],\varnothing\big]\rrb +Z\lrb
\big[\varnothing,[1]\big],\big[\varnothing,[1]\big ] \rrb +Z\lrb \big [
[1],\varnothing\big ],\big[\varnothing,[1]\big ] \rrb +\\ +\nu_1\lrb Z\lrb
[\varnothing,\varnothing],\big [ [1],\varnothing\big]\rrb +Z\lrb
[\varnothing,\varnothing ],\big[\varnothing,[1]\big ] \rrb \rrb
+\nu_2\lrb Z\lrb \big [ [1],\varnothing \big],[\varnothing,\varnothing]\rrb
+Z\lrb \big[\varnothing,[1]\big],[\varnothing,\varnothing]\rrb
\rrb+\nu_1\nu_2+\nu_3
}
\item Conformal block
\eq{5cb11}
{
\mathcal{B}^{(1,1)}=\bg{\alpha_0\beta_0}{\beta_1}(\varnothing,\varnothing,[1])D_
{\beta_1}([1],[1])
\bg{\alpha_1\beta_1}{\beta_2}(\varnothing,[1],[1])D_{\beta_2}([1],[1])
\bg{\alpha_2\beta_2}{\beta_3}(\varnothing,[1],\varnothing)
}
\end{itemize}
\item Expressions for order 3, {\it id est} levels {\bf [3,0]},{\bf [0,3]},{\bf
[2,1]},{\bf [1,2]} are very cumbersome to be presented here.
\end{itemize}
More explicit formulae could be found in the Appendix.
In order to solve the system
\eq{5sys}
{
\left\{\begin{array}{l}
\mathcal{B}^{(1,0)}=\mathcal{Z}^{(1,0)}\\
\mathcal{B}^{(0,1)}=\mathcal{Z}^{(0,1)}\\
\mathcal{B}^{(1,1)}=\mathcal{Z}^{(1,1)}\\
\mathcal{B}^{(2,0)}=\mathcal{Z}^{(2,0)}\\
\mathcal{B}^{(0,2)}=\mathcal{Z}^{(0,2)}
\end{array}\right.,
}
one should use other variables $\sigma_i,\tau_i$
\al
{
\sigma_1&=\mu_1+\mu_2,&\sigma_2&=\mu_1\mu_2\notag\\
\tau_1&=\mu_3+\mu_4, &\tau_2&=\mu_3\mu_4.
}
The relation between the masses in 4d theory and the $\alpha$ parameters in 2d
CFT on sphere is
\al
{
 \mu_1&=-\frac{\epsilon}{2} +\alpha_0+\beta_0,&\mu_2&=\frac{\epsilon}{2}
+\alpha_0-\beta_0\notag\\
 \mu_3&=\frac{3\epsilon}{2} -\alpha_2-\beta_3,&\mu_4&=\frac{\epsilon}{2}
-\alpha_2+\beta_3
&m_1&=\alpha_1,\\
 \nu_1&=\frac{2\alpha_0(\epsilon-\alpha_1)}{\epsilon_1\epsilon_2},
 &\nu_2&=\frac{2\alpha_1(\epsilon-\alpha_2)}{\epsilon_1\epsilon_2},
 &\nu_3&=\frac{2\alpha_0(\epsilon-\alpha_2)}{\epsilon_1\epsilon_2}.\notag
}
together with other solutions, obtained by the
symmetry $\alpha_i\longrightarrow Q-\alpha_i$ for each
$\alpha_i,\beta_0,\beta_3$. These solutions obey all other equations from order two and three.
\subsection{The six-point case}
The $U(1)-$factor is of the form
$(1-x)^{-\nu_1}(1-y)^{-\nu_2}(1-xy)^{-\nu_3}(1-z)^{-\nu_4}(1-yz)^{-\nu_5}
(1-xyz)^{-\nu_6}.$ All the expressions for expansion coefficients could be
straightforwardly written down. However,they are quite involved, and we write them only
symbolically (\rf{ordr1}-\rf{ordr3})
\begin{itemize}
 \item Order {\bf 1}
\begin{figure}[h]
\begin{picture}(450,50)(-90,0)
\put(-10,20){\line(1,0){80}}
\multiput(10,20)(40,0){2}{\line(0,1){20}}
\put(-15,10){$\beta_{0}$}
\put(25,10){$\beta_1$}
\put(60,10){$\beta_2$}
\put(0,45){$\alpha_0$}
\put(45,45){$\alpha_1$}
\put(0,-10){Level {\bf [1,0,0]}}

\put(120,20){\line(1,0){80}}
\multiput(140,20)(40,0){2}{\line(0,1){20}}
\put(125,10){$\beta_{1}$}
\put(155,10){$\beta_2$}
\put(190,10){$\beta_3$}
\put(130,45){$\alpha_1$}
\put(175,45){$\alpha_2$}
\put(135,-10){Level {\bf [0,1,0]}}

\put(250,20){\line(1,0){80}}
\multiput(270,20)(40,0){2}{\line(0,1){20}}
\put(255,10){$\beta_{2}$}
\put(285,10){$\beta_3$}
\put(320,10){$\beta_4$}
\put(260,45){$\alpha_2$}
\put(305,45){$\alpha_3$}
\put(265,-10){Level {\bf [0,0,1]}}
\end{picture}
\caption{\label{fig:ordr1} }
\end{figure}

\item Order {\bf 2}
\begin{figure}[h]
\begin{picture}(450,50)(-40,0)
\put(-10,20){\line(1,0){80}}
\multiput(10,20)(40,0){2}{\line(0,1){20}}
\put(-40,0){Levels {\bf [2,0,0], [0,2,0], [0,0,2]}}

\put(100,20){\line(1,0){80}}
\multiput(120,20)(40,0){2}{\line(0,1){20}}
\put(150,0){Level {\bf [1,0,1]}}
\put(185,30){$\bigotimes$}
\put(200,20){\line(1,0){80}}
\multiput(220,20)(40,0){2}{\line(0,1){20}}
\put(305,20){\line(1,0){120}}
\multiput(325,20)(40,0){3}{\line(0,1){20}}
\put(315,0){Levels {\bf [0,1,1], [1,1,0]}}
\end{picture}
\caption{\label{fig:ordr2} }
\end{figure}

\item Order {\bf 3}
\begin{figure}[h]
\begin{picture}(450,70)(-25,-30)
\put(-15,20){\line(1,0){60}}
\multiput(5,20)(20,0){2}{\line(0,1){20}}
\put(10,0){{\bf [3,0,0]}}
\put(-20,-15){Levels {\bf [0,3,0]}}
\put(10,-30){{\bf [0,0,3]}}
\put(70,20){\line(1,0){60}}
\multiput(90,20)(20,0){2}{\line(0,1){20}}
\put(100,0){Levels {\bf [2,0,1]}}
\put(130,-15){{\bf [1,0,2]}}
\put(135,30){$\bigotimes$}
\put(150,20){\line(1,0){60}}
\multiput(170,20)(20,0){2}{\line(0,1){20}}
\put(245,20){\line(1,0){80}}
\multiput(265,20)(20,0){3}{\line(0,1){20}}
\put(280,0){{\bf [2,1,0]}}
\put(250,-15){Levels {\bf [1,2,0]}}
\put(280,-30){{\bf [0,1,2]}}
\put(280,-45){{\bf [0,2,1]}}
\put(355,20){\line(1,0){100}}
\multiput(375,20)(20,0){4}{\line(0,1){20}}
\put(375,0){Level {\bf [1,1,1]}}
\end{picture}
\caption{\label{fig:ordr3} }
\end{figure}

\end{itemize}
More explicit formulae could be found in the Appendix.
In order to solve the system
\eq{6sys}
{
\left\{\begin{array}{l}
\mathcal{B}^{(1,0,0)}=\mathcal{Z}^{(1,0,0)}\\
\mathcal{B}^{(0,1,0)}=\mathcal{Z}^{(0,1,0)}\\
\mathcal{B}^{(0,0,1)}=\mathcal{Z}^{(0,0,1)}\\
\mathcal{B}^{(1,1,0)}=\mathcal{Z}^{(1,1,0)}\\
\mathcal{B}^{(0,1,1)}=\mathcal{Z}^{(0,1,1)}\\
\mathcal{B}^{(1,1,1)}=\mathcal{Z}^{(1,1,1)}
\end{array}\right.,
}
one should use the variables $\sigma_i,\tau_i$
\al
{
\sigma_1&=\mu_1+\mu_2,&\sigma_2&=\mu_1\mu_2\notag\\
\tau_1&=\mu_3+\mu_4, &\tau_2&=\mu_3\mu_4.
}
The relations between the masses in 4d theory and the $\alpha$ parameters in 2d
CFT on sphere are
\al
{
 \mu_1&=-\frac{\epsilon}{2} +\alpha_0+\beta_0,&\mu_2&=\frac{\epsilon}{2}
+\alpha_0-\beta_0,&m_1&=\alpha_1,\notag\\
 \mu_3&=\frac{3\epsilon}{2} -\alpha_3-\beta_4,&\mu_4&=\frac{\epsilon}{2}
-\alpha_3+\beta_4
&m_2&=\alpha_2,\\
 \nu_1&=\frac{2\alpha_0(\epsilon-\alpha_1)}{\epsilon_1\epsilon_2},
 &\nu_2&=\frac{2\alpha_1(\epsilon-\alpha_2)}{\epsilon_1\epsilon_2},
 &\nu_3&=\frac{2\alpha_0(\epsilon-\alpha_2)}{\epsilon_1\epsilon_2},\notag\\
 \nu_4&=\frac{2\alpha_2(\epsilon-\alpha_3)}{\epsilon_1\epsilon_2},
 &\nu_5&=\frac{2\alpha_1(\epsilon-\alpha_3)}{\epsilon_1\epsilon_2},
 &\nu_6&=\frac{2\alpha_0(\epsilon-\alpha_3)}{\epsilon_1\epsilon_2}.\notag
}
together with the other solutions, obtained by the
symmetries $\alpha_i\longrightarrow Q-\alpha_i$ for each
$\alpha_i,\beta_0,\beta_4$.
These solutions satisfy all other equations in the second and the third orders.

\subsection{The $n-$point case}

In the general case, the relation between masses and $\alpha_i$ was conjectured in \cite{agt}.
After the $5-,6-$point calculations one can reveal a remarkable consequence of the formulae
for the $n-$point conformal block \re{ncb} and the corresponding partition
function \re{nek1}. In the $n-$point case, the conformal block at the levels up to
$(n-4)-th$ reduces to the $4,\dots,(n-1)-$point cases.
This means that the AGT relation obtained from each orders $1,\dots,n-4$ for any
$n$-point conformal block has the same form as the
AGT relations in the $4,\dots,(n-1)-$point cases or the former relations are products of
the latter ones. Consequently,
these relations are satisfied by the same solutions. The first non-trivial
case arises from the level $[1^{n-3}]$. Thus, calculating the first $n-3$ order of the
$4,\dots,n$-point conformal block guarantees the self-consistency of AGT conjecture for the
first $n-3$ orders of the $n$-point case.  The first level always reduces to the $4-$point case, the
second one always reduces to the $4-,5-$point cases and so on because
\al
{
Z_{bifund}(\vec{a},\vec{Y},0,\varnothing,m)=\lrb
Z_{fund}(\vec{a},\vec{Y},\epsilon-m)\rrb^2\label{lft}\\
Z_{bifund}(0,\varnothing,\vec{a},\vec{Y},m)=\lrb Z_{fund}(\vec{a},\vec{Y},m)\rrb^2
\label{rht}.
}
\begin{figure}[h]
%\begin{wrapfigure}{r}{230pt}
\begin{picture}(230,60)(-170,-10)
\put(-10,20){\line(1,0){80}}
\multiput(10,20)(40,0){2}{\line(0,1){20}}
\put(-15,10){$\beta_{i-1}$}
\put(25,10){$\beta_i$}
\put(60,10){$\beta_{i+1}$}
\put(0,45){$\alpha_{i-1}$}
\put(45,45){$\alpha_{i}$}
\put(25,-10){$a)$}

\put(110,20){\line(1,0){80}}
\multiput(130,20)(40,0){2}{\line(0,1){20}}
\put(105,10){$\beta_{i}$}
\put(140,10){$\beta_{i+1}$}
\put(180,10){$\beta_{i+2}$}
\put(125,45){$\alpha_{i}$}
\put(165,45){$\alpha_{i+1}$}
\put(145,-10){$b)$}
\end{picture}
\caption{\label{fig:fig1l}}
%\end{wrapfigure}
\end{figure}
Now let us discuss general relations between the masses and
conformal dimensions. For instance, for any $n-$point case the first order AGT relations reduces to the $4-$point case
\al
{
\mathcal{B}&=\sum\limits_{|Y_i|=l_i}x^{l_i}\sum\limits_{Y_i'}\bg{\alpha_{i-1}\beta_{i-1}}
{\beta_i}(\varnothing,\varnothing,Y_i')D_{\beta_i}(Y_i',Y_i)\bg{\alpha_i\beta_i}{\beta_
{i+1}}(\varnothing,Y_{i},\varnothing),\\
\mathcal{Z}&=Z^{U(1)}Z_{bifund}(0,\varnothing,\vec{a}_i,\vec{Y}_i,m_{i-1})Z_{vector}
(\vec{a}_i,\vec{Y}_i)Z_{bifund}(\vec{a}_i,\vec{Y}_i,0,\varnothing,m_i),
}
these formulae correspond to diagram \rf{fig1l}$a)$. The second order in the $5-$case reduces to the $4-$point case and contains one new level {\bf [1,1]}. The second order in the $6-$point case reduces to the $4-$point case (levels {\bf [2,0,0],[0,2,0],[0,0,2]}) \re{4cb}, to product of the $4-$point cases ( level {\bf [1,0,1]}) and to the $5-$point case (levels {\bf [0,1,1],[1,1,0]}). The second order of the $6-$point case contains one non-trivial level {\bf [1,1,1]}. All other levels reduce to the $4-point$, product of two $4-$point and to the $5-$point cases.

To obtain formulae for $m_i$ one  should consider two series of the levels
\par
$[0,\dots,0,l_i,0,\dots,0], l_i \in \mathds{N}$ and
$[0,\dots,0,l_{i+1},0,\dots,0], l_{i+1} \in \mathds{N}$
\al
{
\mathcal{B}&=\sum\limits_{|Y_i|=l_i}x^{l_i}\sum\limits_{Y_i'}\bg{\alpha_{i-1}\beta_{i-1}}
{\beta_i}(\varnothing,\varnothing,Y_i')D_{\beta_i}(Y_i',Y_i)\bg{\alpha_i\beta_i}{\beta_
{i+1}}(\varnothing,Y_{i},\varnothing),\\
\mathcal{Z}&=Z^{U(1)}Z_{bifund}(0,\varnothing,\vec{a}_i,\vec{Y}_i,m_{i-1})Z_{vector}
(\vec{a}_i,\vec{Y}_i)Z_{bifund}(\vec{a}_i,\vec{Y}_i,0,\varnothing,m_i),
} see diagram \rf{fig1l}$a)$
\al
{
\mathcal{B}&=\sum\limits_{|Y_{i+1}|=l_{i+1}}x^{l_{i+1}}\sum\limits_{Y_i'}\bg{
\alpha{i}\beta_{1}}{\beta_{i+1}}(\varnothing,\varnothing,Y_{i+1}')D_{\beta_{i+1}}(Y_{i+1
}',Y_{i+1})\bg{\alpha{i+1}\beta_{i+1}}{\beta_{i+2}}(\varnothing,Y_{i+1},\varnothing),\\
\mathcal{Z}&=Z^{U(1)}Z_{bifund}(0,\varnothing,\vec{a}_{i+1},\vec{Y}_{i+1},m_{i})Z_{
vector}(\vec{a}_{i+1},\vec{Y}_{i+1})Z_{bifund}(\vec{a}_{i+1},\vec{Y}_{i+1},0,
\varnothing,m_{i+1}),
}
and these formulae correspond to diagram \rf{fig1l}$b)$.
Solving these two systems one obtains two sets of solution for $m_i$ from each system.
So the general solution for $n$-point which reduces to these two systems will be
\al
{ \mu_1&=-\frac{\epsilon}{2} +\alpha_0+\beta_0,&\mu_2&=\frac{\epsilon}{2}
+\alpha_0-\beta_0,&m_i&=\alpha_i,\notag\\
 \mu_3&=\frac{3\epsilon}{2} -\alpha_{n-3}-\beta_{n-2},&\mu_4&=\frac{\epsilon}{2}
-\alpha_{n-3}+\beta_{n-2},\label{e:glans}\\
 \nu_{ij}&=\frac{2\alpha_i(\epsilon-\alpha_j)}{\epsilon_1\epsilon_2}.\notag
}
or any other solutions, made from this one using general symmetries, described in Section \ref{symmsec}.

\section{Conclusion}
We checked the AGT conjecture up to the third level for the 4-,5-,6-points conformal blocks and confirmed all the predictions. We also showed that correct universal formulae for parameters of $U(1)$ factors are given by \re{glans}.
Also we discuss the possible expression for $U(1)-$factor. The exact proof of the AGT relations within this framework requires a manifest expression for the conformal block. However, it is not available, since no exact general formulas both for vertices and the Shapovalov form are known. The non-comb type conformal blocks (non-linear quiver theories case) require more complicated generalization of the AGT framework.

\section*{Acknowledgements}
We appreciate very useful discussions with Andrei Mironov, Alexei Morozov. We are also grateful to Shamil Shakirov for the help with computer calculations. V.Alba appreciates very useful discussions with Alexander Belavin and Yaroslav Pugai. We are grateful to Alexander Belavin for his excellent lectures on CFT.

V. Alba's research was held within the bounds of Federal Program "Scientific and Scientific-Pedagogical personnel of innovational Russia" and was partly supported by RFBR grant 10-02-00499. The work of And. Morozov was partly supported by grants RFBR 07-01-00526 and NSh-3036.2008.2.

\newpage
\appendix
\section{List of vertices}
In this appendix the list of different vertices used in
this paper is given. Useful formula
\eq{mf}
{
\boxed{
\bg{23}{1}([Y_2,1],[Y_3],[Y_1])=\lrb\dl{1}-\dl{2}-\dl{3}
+|Y_1|-|Y_2|-|Y_3|\rrb\bg{23}{1}([Y_2],[Y_3],[Y_1]).
}
}
The full list of the vertices (most of them require direct calculation without \re{mf})

\al
{
%\bg{23}{1}(Y_2,Y_3,Y_1)&\equiv\ltb L_{-Y_1} V_1\big| L_{-Y_2}V_2(1)L_{-Y_3}V_3(0)\rtb,
%\\
\bg{23}{1}([1],\varnothing,\varnothing)&=
-\bg{23}{1}(\varnothing,[1],\varnothing),
\\
%&\ltb V_1\big| (L_{-1}V_2)(1) V_3(0)\rtb \propto
\bg{23}{1}([1],\varnothing,\varnothing)&= (\Delta_1-\Delta_2-\Delta_3),
\\
%&\ltb V_1\big| V_2(1)(L_{-1} V_3)(0)\rtb \propto
\bg{23}{1}(\varnothing,[1],\varnothing)&=(\Delta_3+\Delta_2-\Delta_1),
\\
%&\ltb L_{-1}V_1\big| V_2(1) V_3(0)\rtb \propto
\bg{23}{1}(\varnothing,\varnothing,[1])&= (\Delta_1+\Delta_2-\Delta_3),
\\
%&\ltb V_1\big| L_{-1}V_2(1) L_{-1}V_3(0)\rtb \propto
\bg{23}{1}([1],[1],\varnothing)&=
-(\Delta_3+\Delta_2-\Delta_1)(\Delta_3+\Delta_2-\Delta_1+1),
\\
%&\ltb L_{-1}V_1\big| V_2(1) L_{-1}V_3(0)\rtb \propto
\bg{23}{1}(\varnothing,[1],[1])&=
(\Delta_3+\Delta_2-\Delta_1)(\Delta_1+\Delta_2-\Delta_3-1)+2\Delta_3,
\\
%&\ltb L_{-1}V_1\big| L_{-1}V_2(1) V_3(0)\rtb \propto
\bg{23}{1}([1],\varnothing,[1])&=
(\Delta_1+\Delta_2-\Delta_3)(\Delta_1-\Delta_2-\Delta_3+1),
\\
%&\ltb V_1\big| L_{-1}^2V_2(1) V_3(0)\rtb \propto
\bg{23}{1}([1^2],\varnothing,\varnothing)&=
(\Delta_1-\Delta_2-\Delta_3)(\Delta_1-\Delta_2-\Delta_3-1),
\\
%&\ltb V_1\big| V_2(1) L_{-1}^2V_3(0)\rtb \propto
\bg{23}{1}(\varnothing,[1^2],\varnothing)&=
(\Delta_1-\Delta_2-\Delta_3)(\Delta_1-\Delta_2-\Delta_3-1),
\\
%&\ltb L_{-1}^2 V_1\big| V_2(1) V_3(0)\rtb \propto
\bg{23}{1}(\varnothing,\varnothing,[1^2])&=
(\Delta_1+\Delta_2-\Delta_3)(\Delta_1+\Delta_2-\Delta_3+1),
\\
%&\ltb L_{-2} V_1\big| V_2(1) V_3(0)\rtb \propto
\bg{23}{1}(\varnothing,\varnothing,[2])&= (\Delta_1+2\Delta_2-\Delta_3),
\\
%&\ltb V_1\big| V_2(1) L_{-2}V_3(0)\rtb \propto
\bg{23}{1}(\varnothing,[2],\varnothing)&=
(\Delta_3+2\Delta_2-\Delta_1),
\\
%&\ltb V_1\big| L_{-1}V_2(1) L_{-2}V_3(0)\rtb \propto
\bg{23}{1}([1],[2],\varnothing)&=
(\Delta_1-\Delta_2-\Delta_3-2)(\Delta_3+2\Delta_2-\Delta_1),
\\
%&\ltb L_{-2}V_1\big| V_2(1) L_{-1}V_3(0)\rtb \propto
\bg{23}{1}(\varnothing,[1],[2])&=
(\Delta_3+\Delta_2-\Delta_1)(\Delta_1+2\Delta_2-\Delta_3-1),
\\
%&\ltb L_{-1}V_1\big| V_2(1) L_{-2}V_3(0)\rtb \propto
\bg{23}{1}(\varnothing,[2],[1])&=
(\Delta_3+2\Delta_2-\Delta_1)(\Delta_1+\Delta_2-\Delta_3-2)+
3(\Delta_3+\Delta_2-\Delta_1),
\\
%&\ltb L_{-1}^2V_1\big| V_2(1) L_{-1}V_3(0)\rtb \propto
\bg{23}{1}(\varnothing,[1],[1^2])&= 2\Delta_3(\Delta_1+\Delta_2-\Delta_3)
+2\Delta_3(\Delta_1+\Delta_2-\Delta_3-1)+\notag\\
&\quad+(\Delta_3+\Delta_2-\Delta_1)(\Delta_1+\Delta_2-\Delta_3-1)^2,
\\
%&\ltb L_{-1}V_1\big| V_2(1) L_{-1}^2V_3(0)\rtb \propto
\bg{23}{1}(\varnothing,[1^2],[1])&= 2(2\dl{3}+1)(\dl{3}+\dl{2}-\dl{1})+\notag\\
&\quad+(\Delta_1-\Delta_2-\Delta_3)(\Delta_1-\Delta_2-\dl{3}
-1)(\Delta_1+\Delta_2-\Delta_3-2),
\\
%&\ltb L_{-1}V_1\big|L_{-1} V_2(1) L_{-1}V_3(0)\rtb \propto
\bg{23}{1}([1],[1],[1])&=
2\dl{3}(\dl{1}-\dl{2}-\dl{3}-1)
+(\Delta_1-\Delta_2-\Delta_3)(\Delta_3+\Delta_2-\Delta_1)
(\Delta_1+\Delta_2-\Delta_3-1),\\
%&\ltb L_{-1}^2 V_1\big| L_{-1}V_2(1) V_3(0)\rtb \propto
\bg{23}{1}([1],\varnothing,[1^2])&=
(\dl{1}-\dl{2}-\dl{3}+1)(\dl{1}+\dl{2}-\dl{3})(\dl{1}+\dl{2}-\dl{3}+1),
\\
%&\ltb V_1\big| L_{-1}V_2(1) L_{-1}^2V_3(0)\rtb \propto
\bg{23}{1}([1],[1^2],\varnothing)&=
(\Delta_1-\Delta_2-\Delta_3)(\Delta_1-\Delta_2-\Delta_3-1)
(\Delta_1-\Delta_2-\Delta_3-2),
\\
%&\ltb L_{-1} V_1\big| L_{-1}^2V_2(1) V_3(0)\rtb \propto
\bg{23}{1}([1^2],\varnothing,[1])&=
(\Delta_1-\Delta_2-\Delta_3)(\Delta_1-\Delta_2-\Delta_3-1)
(\Delta_1+\Delta_2-\Delta_3),
\\
%&\ltb L_{-2}V_1\big| V_2(1) L_{-2}V_3(0)\rtb \propto
\underline{\bg{23}{1}(\varnothing,[2],[2])}&=
(\Delta_3+2\Delta_2-\Delta_1)(\Delta_1+2\Delta_2-\Delta_3-2)+4\Delta_3+
\frac{c}{2},
\\
%&\ltb L_{-2}V_1\big| V_2(1) L_{-1}^2V_3(0)\rtb \propto
\underline{\bg{23}{1}(\varnothing,[1^2],[2])}&=
6\Delta_3+(\Delta_1-\Delta_2-\Delta_3)(\Delta_1-\Delta_2-\Delta_3-1)
(\Delta_1+2\Delta_2-\Delta_3-2),
\\
%&\ltb L_{-1}V_1\big| L_{-1}V_2(1) L_{-2}V_3(0)\rtb \propto
\bg{23}{1}([1],[2],[1])&=
(\Delta_3+2\Delta_2-\Delta_1)(\Delta_1+\Delta_2-\Delta_3-2)
(\Delta_1-\Delta_2-\Delta_3-1)+\notag\\
&\quad+3(\Delta_3+\Delta_2-\Delta_1)(\Delta_1-\Delta_2-\Delta_3-1),
\\
%&\ltb V_1\big| L_{-1}^2V_2(1) L_{-2}V_3(0)\rtb \propto
\bg{23}{1}([1^2],[2],\varnothing)&=
(\Delta_1-\Delta_2-\Delta_3-3)(\Delta_1-\Delta_2-\Delta_3-2)
(\Delta_3+2\Delta_2-\Delta_1),
\\
%&\ltb L_{-1}^2V_1\big| V_2(1) L_{-2}V_3(0)\rtb \propto
\underline{\bg{23}{1}(\varnothing,[2],[1^2])}&=
6(\Delta_1+\Delta_2-\Delta_3-1)(\Delta_3+\Delta_2-\Delta_1)+\notag\\
&\quad+(\Delta_1+\Delta_2-\Delta_3-1)(\Delta_3+2\Delta_2-\Delta_1)
(\Delta_1+\Delta_2-\Delta_3-2)+6\dl{3},
\\
%&\ltb V_1\big| L_{-1}^2V_2(1) L_{-1}^2V_3(0)\rtb \propto
\bg{23}{1}([1^2],[1^2],\varnothing)&=
(\Delta_1-\Delta_2-\Delta_3)(\Delta_1-\Delta_2-\Delta_3-1)\times\\
&\quad\times(\Delta_1-\Delta_2-\Delta_3-2)(\Delta_1-\Delta_2-\Delta_3-3),
\\
%&\ltb V_1\big| L_{-1}V_2(1) L_{-1}^3V_3(0)\rtb \propto
\bg{23}{1}([1],[1^3],\varnothing)&=
-(\Delta_1-\Delta_2-\Delta_3)(\Delta_1-\Delta_2-\Delta_3-1)
\times(\Delta_1-\Delta_2-\Delta_3-2)(\Delta_1-\Delta_2-\Delta_3-3),
\\
%&\ltb L_{-1}V_1\big| L_{-1}V_2(1) L_{-1}^2V_3(0)\rtb \propto
\bg{23}{1}([1],[1^2],[1])&=
2(2\dl{3}+1)(\dl{3}+\dl{2}-\dl{1})(\dl{1}-\dl{2}-\dl{3}-1)+\notag\\
&\quad+(\Delta_1-\Delta_2-\Delta_3)(\Delta_1-\Delta_2-\dl{3}-1)^2
(\Delta_1+\Delta_2-\Delta_3-2),
\\
%&\ltb L_{-1}^2V_1\big| V_2(1) L_{-1}^2V_3(0)\rtb \propto
\underline{\bg{23}{1}(\varnothing,[1^2],[1^2])}&=
2\Delta_2\bg{23}{1}(\varnothing,[1^2],[1])
+\bg{23}{1}([1],[1^2],[1])
+2(2\Delta_3+1)\bg{23}{1}(\varnothing,[1],[1])
}

\section{Explicit evaluation}
\subsection{The four-point conformal block and $U(2)$-quiver}

\begin{itemize}
\item \textbf{Level {\bf 1}}
 \begin{align}
\hline
&&\mbox{Conformal block}&\displaystyle =\quad \mbox{Nekrasov's partition function}\notag\\
\hline
&&&\notag\\
a^4&& -2&=-2 \label{cb4g1l4}\\
a^2&& \epsilon^2-2(\alpha_{0}^2-\beta_{0}^2+\alpha_{1}^2-\beta_{2}^2)-2\epsilon(\alpha_{0}-\beta_{0}+\alpha_{1}-\beta_{2})
&=-2 \sigma_2+ \epsilon \sigma_1+4\nu\epsilon_1\epsilon_2\label{cb4g1l2}\\
a^0&& -2\lrb\epsilon^2/2+\epsilon(\alpha_{0}-\beta_{0})-(\alpha^2_1-\beta_{0}^2)\rrb\times&=-2\sigma_4+\epsilon \sigma_3 -\nu\epsilon_1 \epsilon_2\epsilon^2\notag\\
&&\lrb\epsilon^2/2+\epsilon(\alpha_{1}-\beta_{2})-(\alpha^2_3-\beta_{2}^2)\rrb& \label{cb4g1l0}
\end{align}
\item \textbf{Level {\bf 2}}
\begin{align}
\hline
&&\mbox{Conformal block}&\displaystyle =\quad \mbox{Nekrasov's partition function}\notag\\
\hline
&&&\notag\\
a^{10}&& 16&=16\label{cb4g2l10}\\
a^8&&-\lfb 32\epsilon^2 +18\epsilon_1\epsilon_2\right.&=16(2\sigma_2-\epsilon \sigma_1-\epsilon^2)-18\epsilon_1\epsilon_2\notag\\
&&\left.+32\lrb\epsilon(\alpha_{0}-\beta_{0}+\alpha_{1}-\beta_{2})+\alpha_{0}^2-\beta_{0}^2+\alpha_{1}^2-\beta_{2}^2\rrb\rfb &\quad -64\nu\epsilon_1\epsilon_2\label{cbg2l8}\\
&&a^6\dots a^4\dots a^2\dots a^0&\notag\\
\end{align}
\end{itemize}

\subsection{The five-point conformal block and $U(2)\otimes U(2)$-quiver}

\begin{itemize}
 \item Level \textbf{[1,0]}
\begin{align}
 \hline
&&\mbox{Conformal block}\quad&\displaystyle =\quad \mbox{Nekrasov's partition function}\notag\\
\hline
&&&\notag\\
a_1^4a_2^0&&2&\displaystyle =2 \label{cbxy10l40}
\\
a_1^2a_2^2&&-2&\displaystyle =-2 \label{cbxy10l22}
\\
a_1^2a_2^0&&-\epsilon^2/2+2(\alpha_{0}^2+\beta_{0}^2-\alpha_{2}^2)-2\epsilon(\alpha_{0}+\beta_{0}-\alpha_{2})&\displaystyle =2\mu_1\mu_2+(\mu_1+\mu_2)(3\epsilon-4m_1)+\notag\\
&&&\quad+2m_1(m_1-\epsilon) -4\nu_1\epsilon_1\epsilon_2\label{cbxy10l20}
\\
a_1^0a_2^2&&\epsilon^2/2-2(\beta_{0}-\alpha_{2})(\beta_{0}+\alpha_{2}-\epsilon)&\displaystyle =\epsilon(\mu_1+\mu_2)-2\mu_1\mu_2\label{cbxy10l02}
\\
a_1^0a_2^0&&\alpha_{0}(\epsilon-\alpha_{0})(\epsilon^2/2+2(\beta_{0}-\alpha_{2})(\epsilon-(\beta_{0}+\alpha_{2})))&\displaystyle =2m_1\mu_1\mu_2(m_1-\epsilon)-\notag\\
&&&\quad-\epsilon(\mu_1+\mu_2)(m_1-\epsilon)^2+\nu_1\epsilon_1\epsilon_2\epsilon^2\label{cbxy10l00}
\end{align}
\item Level \textbf{[0,1]}
\begin{align}
 \hline
&&\mbox{Conformal block}\quad&\displaystyle =\quad \mbox{Nekrasov's partition function}\notag\\
\hline
&&&\notag\\
a_1^0a_2^4&&2&\displaystyle =2 \label{cb501l40}
\\
a_1^2a_2^2&&-2&\displaystyle =-2 \label{cb501l22}
\\
a_1^0a_2^2&&-\epsilon^2/2+2(\alpha_{0}^2+\beta_{3}^2-\alpha_{1}^2)-2\epsilon(\alpha_{0}+\beta_{3}-\alpha_{1})&\displaystyle =2\mu_3\mu_4+(\mu_3+\mu_4)(4m_1-\epsilon)+\notag\\
&&&\quad+2m_1(m_1-\epsilon)-4\nu_2\epsilon_1\epsilon_2\label{cb501l02}
\\
a_1^2a_2^0&&\epsilon^2/2-2(\beta_{3}-\alpha_{1})(\beta_{3}+\alpha_{1}-\epsilon)&\displaystyle =\epsilon(\mu_3+\mu_4)-2\mu_3\mu_4\label{cb501l20}
\\
a_1^0a_2^0&&\alpha_{0}(\epsilon-\alpha_{0})(\epsilon^2/2+2(\beta_{3}-\alpha_{1})(\epsilon-\beta_{3}-\alpha_{1}))&\displaystyle =2m_1\mu_3\mu_4(m_1-\epsilon)-\notag\\
&&&
\quad-\epsilon(\mu_3+\mu_4)m_1^2+\nu_2\epsilon_1\epsilon_2\epsilon^2\label{cb501l00}
\end{align}
\end{itemize}

\begin{enumerate}
\item
\al
{\sigma_1 &= 2 \beta_{0},
	&\sigma_2 &=\lrb -\frac{\epsilon}{2} + \beta_{0} + \alpha_{2}\rrb
 \lrb\frac{\epsilon}{2} + \beta_{0} -\alpha_{2}\rrb,\notag\\
 \tau_1 &= 2 (\epsilon - \beta_{3}),
	&\tau_2 &= \lrb\frac{\epsilon}{2} - \beta_{3} + \alpha_{1}\rrb
 \lrb \frac{3 \epsilon}{2}- \beta_{3} -\alpha_{1}\rrb,\notag\\
	\nu_1 &=\frac{2 \beta_{0} (\epsilon - \alpha_{0})}{\epsilon_1 \epsilon_2},
	&\nu_2 &=\frac{2 \alpha_{0}(\epsilon - \beta_{3})}{\epsilon_1 \epsilon_2},
	&\nu_3 &=\frac{2 \beta_{0} (\epsilon - \beta_{3})}{\epsilon_1
\epsilon_2},\notag\\
	m_1 &= \alpha_{0}.
}
\item
\al
{
\sigma_1 &= 2 \epsilon_2 + 2 \epsilon_1 - 2 \beta_{0},& \sigma_2 &= 1/4
 (\epsilon - 2 \beta_{0} + 2 \alpha_{2})
 (3 \epsilon_1 + 3 \epsilon_2 - 2 \beta_{0} - 2 \alpha_{2}),\notag\\
 \tau_1 &= 2 \beta_{3},& \tau_2 &= -1/4
 (\epsilon - 2 \beta_{3} - 2 \alpha_{1})
 (\epsilon + 2 \beta_{3} - 2 \alpha_{1}),\notag\\
 \nu_1 &=\frac{2 \alpha_{0} (\epsilon - \beta_{0})}{\epsilon_1 \epsilon_2},
 &\nu_2 &=\frac{2 \beta_{3} (\epsilon - \alpha_{0})}{\epsilon_1 \epsilon_2},
 &\nu_3 &=\frac{2 \beta_{3} (\epsilon - \beta_{0})}{\epsilon_2 \epsilon_1},\notag
\\
 m_1 &= \epsilon - \alpha_{0}.
}
\item
\al
{ \sigma_1 &= 2 \epsilon_2 + 2 \epsilon_1 - 2 \beta_{0},& \sigma_2 &= 1/4
 (\epsilon - 2 \beta_{0}
  + 2 \alpha_{2})
 (3 \epsilon_1 + 3 \epsilon_2 - 2 \beta_{0} - 2 \alpha_{2}),\notag\\
 \tau_1 &= 2 \beta_{3},& \tau_2 &= -1/4
 (\epsilon - 2 \beta_{3} - 2 \alpha_{1})
 (\epsilon + 2 \beta_{3} - 2 \alpha_{1}),\notag\\
 \nu_1 &=\frac{2 (\epsilon - \beta_{0}) (\epsilon - \alpha_{0})}{\epsilon_1
\epsilon_2},
 & \nu_2 &=\frac{2 \beta_{3} \alpha_{0}}{\epsilon_1 \epsilon_2},
 & \nu_3 &=\frac{2 \beta_{3} (\epsilon - \beta_{0})}{\epsilon_2
\epsilon_1},\notag\\
 m_1 &= \alpha_{0}.
}

\item
\al
{
\sigma_1 &= 2 \beta_{0}, &\sigma_2 &= -1/4 (\epsilon - 2 \beta_{0}
 - 2 \alpha_{2})
 (\epsilon + 2 \beta_{0} - 2 \alpha_{2}),\notag\\
	\tau_1 &= 2 \beta_{3}, &\tau_2 &= -1/4 (\epsilon - 2 \beta_{3} - 2
\alpha_{1})
 (\epsilon + 2 \beta_{3} - 2 \alpha_{1}),\notag\\
 \nu_1&=\frac{2 \beta_{0} (\epsilon - \alpha_{0})}{\epsilon_1 \epsilon_2},
 &\nu_2 &=\frac{2 \beta_{3} \alpha_{0}}{\epsilon_1 \epsilon_2}, &\nu_3 &= \frac{2
\beta_{0} \beta_{3}}{\epsilon_1 \epsilon_2},\notag\\
 m_1 &= \alpha_{0}.
}
\item
\al
{ \sigma_1 &= 2 \beta_{0}, &\sigma_2 &= -1/4 (\epsilon - 2 \beta_{0} - 2 \alpha_{2})
 (\epsilon + 2 \beta_{0} - 2 \alpha_{2}),\notag\\
 \tau_1 &= 2 \beta_{3}, &\tau_2 &= -1/4 (\epsilon - 2 \beta_{3} - 2 \alpha_{1})
 (\epsilon + 2 \beta_{3} - 2 \alpha_{1}),\notag\\
 \nu_1 &=\frac{2\alpha_1 \beta_{0}}{\epsilon_1 \epsilon_2},
 &\nu_2 &=\frac{2\beta_{3} (\epsilon - \alpha_{0})}{\epsilon_1 \epsilon_2},
 &\nu_3 &=\frac{2 \beta_{0} \beta_{3}}{\epsilon_1 \epsilon_2},\notag\\
 m_1 &= \epsilon - \alpha_{0}.
}

\item
\al
{ \sigma_1 &= 2 \epsilon_2 + 2 \epsilon_1 - 2 \beta_{0}, &\sigma_2 &= 1/4
 (\epsilon - 2 \beta_{0} + 2 \alpha_{2})(3 \epsilon_1 + 3 \epsilon_2 - 2 \beta_{0} -
2 \alpha_{2}),\notag\\
 \tau_1 &= 2 \epsilon_1 + 2 \epsilon_2 - 2 \beta_{3}, &\tau_2 &= 1/4
 (\epsilon - 2 \beta_{3} + 2 \alpha_{1})(3 \epsilon_1 + 3 \epsilon_2 - 2 \beta_{3} -
2 \alpha_{1}),\notag\\
 \nu_1 &=\frac{2 (\epsilon - \beta_{0}) (\epsilon - \alpha_{0})}{\epsilon_1
\epsilon_2},
 &\nu_2 &=\frac{2 \alpha_{0}(\epsilon - \beta_{3})}{\epsilon_1 \epsilon_2},
 &\nu_3 &=\frac{2 (\epsilon - \beta_{3}) (\epsilon - \beta_{0})}{\epsilon_1
\epsilon_2},\notag\\
 m_1 &= \alpha_{0}.
}
\item
\al
{ \sigma_1 &= 2 \beta_{0}, &\sigma_2 &= -1/4(\epsilon - 2 \beta_{0} - 2 \alpha_{2})
 (\epsilon + 2 \beta_{0} - 2 \alpha_{2}),\notag\\
 \tau_1 &= 2 \epsilon_1 + 2 \epsilon_2 - 2 \beta_{3}, &\tau_2 &= 1/4
 (\epsilon - 2 \beta_{3} + 2 \alpha_{1})(3 \epsilon_1 + 3 \epsilon_2 - 2 \beta_{3} -
2 \alpha_{1}),\notag\\
 \nu_1 &=\frac{2 \alpha_{0} \beta_{0}}{\epsilon_1 \epsilon_2},
 &\nu_2 &=\frac{2 (\epsilon - \beta_{3}) (\epsilon - \alpha_{0})}{\epsilon_1
\epsilon_2},
 &\nu_3 &=\frac{2 \beta_{0} (\epsilon - \beta_{3})}{\epsilon_1
\epsilon_2},\notag\\
 m_1 &= \epsilon - \alpha_{0}
}
\item
\al
{ \sigma_1 &= 2 \epsilon_2 + 2 \epsilon_1 - 2 \beta_{0}, &\sigma_2 &= 1/4
 (\epsilon - 2 \beta_{0} + 2 \alpha_{2})(3 \epsilon_1 + 3 \epsilon_2 - 2 \beta_{0} -
2 \alpha_{2}),\notag\\
 \tau_1 &= 2 \epsilon_1 + 2 \epsilon_2 - 2 \beta_{3}, &\tau_2 &= 1/4
 (\epsilon - 2 \beta_{3} + 2 \alpha_{1})(3 \epsilon_1 + 3 \epsilon_2 - 2 \beta_{3} -
2 \alpha_{1}),\notag\\
 \nu_1 &=\frac{2 \alpha_{0} (\epsilon - \beta_{0})}{\epsilon_1 \epsilon_2},
 &\nu_2 &=\frac{2 (\epsilon - \beta_{3}) (\epsilon - \alpha_{0})}{\epsilon_1
\epsilon_2},
 &\nu_3 &=\frac{2 (\epsilon - \beta_{3}) (\epsilon - \beta_{0})}{\epsilon_1
\epsilon_2},\notag\\
 m_1 &= \epsilon - \alpha_{0}.
}
\end{enumerate}
\subsection{The six-point conformal block and
$U(2)\otimes U(2)\otimes U(2)$-quiver}
\begin{enumerate}
 \item Level \textbf{[1,0,0]}
\begin{align}
 \hline
&&\mbox{Conformal block}\quad&\displaystyle =\quad \mbox{Nekrasov's partition function}\notag\\
\hline
&&&\notag\\
a_1^4a_2^0a_3^0&&2&\displaystyle =2
\\
a_1^2a_2^2a_3^0&&-2&\displaystyle =-2
\\
a_1^2a_2^0a_3^0&&\epsilon^2/2-2\epsilon(\alpha_0+\alpha_1-\beta_0)+2\alpha_1^2-2\beta_0^2&\displaystyle =2m_1^2-4\nu_1\epsilon_1\epsilon_2-2\epsilon_1m_1-2\epsilon_2m_1+\notag\\
&&&\quad+(3\epsilon_2-4m_1+3\epsilon_1)\sigma_1+2\sigma_2\notag\\
a_1^0a_2^2a_3^0&&\epsilon^2/2+2(\alpha_0-\beta_0)(\epsilon-\alpha_0-\beta_0)&=
\epsilon\sigma_1-2\sigma_2\notag\\
a_1^0a_2^0a_3^0&&1/2\alpha_1(\epsilon-\alpha_1)\lrb\epsilon^2-4\epsilon(\beta_0-\alpha_0)-4(\alpha_0^2+\beta_0^2)\rrb&=\nu_1\epsilon_1\epsilon_2^3+\nu_1\epsilon_1^3\epsilon_2+2\nu_1\epsilon_1^2\epsilon_2^2+\notag\\
&&&\quad+(-\epsilon_2^3-\epsilon_1^3-m_1^2\epsilon_2-3\epsilon_1\epsilon_2^2+2\epsilon_1^2m_1-m_1^2\epsilon_1+\notag\\
&&&\quad+2\epsilon_2^2m_1-3\epsilon_1^2\epsilon_2+4\epsilon_1m_1\epsilon_2)\sigma_1+\notag\\
&&&\quad+(-2\epsilon_2m_1-2\epsilon_1m_1+2m_1^2)\sigma_2
\end{align}

\item Level{\bf [0,1,0]}
\begin{align}
 \hline
&&\mbox{Conformal block}\quad&\displaystyle =\quad \mbox{Nekrasov's partition function}\notag\\
\hline
&&&\notag\\
a_1^0a_2^4a_3^0&&2&=2
\\
a_1^2a_2^2a_3^0&&-2&=-2
\\
a_1^0a_2^2a_3^2&&-2&=-2
\\
a_1^2a_2^0a_3^2&&2&=2
\\
a_1^2a_2^0a_3^0&&2\alpha_2(\epsilon-\alpha_2)&=2m_2(\epsilon-m_2)\notag\\
a_1^0a_2^2a_3^0&&2(\alpha_1^2+\alpha_2^2)-2\epsilon(\alpha_1+\alpha_2)&=
2m_2^2+2m_1^2-2\epsilon_1m_2-2\epsilon_2m_2+\notag\\
&&&\quad+6\epsilon_1m_1+6\epsilon_2m_1-8m_1m_2-4\nu_2\epsilon_1\epsilon_2\notag\\
a_1^0a_2^0a_3^2&&2\alpha_1(\epsilon-\alpha_1)&=-2m_1^2+2\epsilon_1m_1+2\epsilon_2m_1\notag\\
a_1^0a_2^0a_3^0&&2\alpha_1\alpha_2(\epsilon-\alpha_1)(\epsilon-\alpha_2)&=8m_1\epsilon_1m_2\epsilon_2+2m_1^2m_2^2-2m_1^2\epsilon_1m_2-2m_1^2\epsilon_2m_2-\notag\\
&&&\quad-6m_1\epsilon_1^2\epsilon_2-6m_1\epsilon_1\epsilon_2^2+4m_1\epsilon_1^2m_2+4m_1\epsilon_2^2m_2-2m_1m_2^2\epsilon_1-\notag\\
&&&\quad-2m_1m_2^2\epsilon_2+\nu_2\epsilon_1^3\epsilon_2+2\nu_2\epsilon_1^2\epsilon_2^2+\nu_2\epsilon_1\epsilon_2^3-2m_1\epsilon_1^3-2m_1\epsilon_2^3
\end{align}

\item Level {\bf [0,0,1]}
\begin{align}
 \hline
&&\mbox{Conformal block}\quad&\displaystyle =\quad \mbox{Nekrasov's partition function}\notag\\
\hline
&&&\notag\\
a_1^0a_2^0a_3^4&&2&=2
\\
a_1^0a_2^2a_3^2&&-2&=-2
\\
a_1^0a_2^2a_3^0&&\epsilon^2/2+2(\alpha_3-\beta_4)(\epsilon-\alpha_3-\beta_4)&=\tau_1\epsilon-2\tau_2
\\
a_1^2a_2^0a_3^2&&-\epsilon^2/2-2\epsilon(\alpha_2+\alpha_3-\beta_4)+2(\alpha_3^2-\beta_4^2)&=-4\nu_3\epsilon_1\epsilon_2-2\epsilon_1m_2-2\epsilon_2m_2+2m_2^2-\tau_1\epsilon_2+\notag\\
&&&\quad+4\tau_1m_2-\tau_1\epsilon_1+2\tau_2
\end{align}

\item Level {\bf [1,1,1]}
\begin{align}
 \hline
&&\mbox{Conformal block}\quad&\displaystyle =\quad \mbox{Nekrasov's partition function}\notag\\
\hline
&&&\notag\\
a_1^6a_2^4a_3^2&&2&=2
\\
a_1^6a_2^4a_3^0&&8\tau_2-4\epsilon\tau_1&=+2\epsilon^2-8(\alpha_3-\beta_4)(\epsilon-\alpha_3-\beta_4)
\\
a_1^6a_2^2a_3^2&&4\epsilon^2+2\epsilon_1\epsilon_2+16(\alpha_3-\beta_4)(\epsilon-\alpha_3-\beta_4)&=8\epsilon_1\epsilon_2+16\nu_3\epsilon_1\epsilon_2-16\tau_1m_2+8\tau_1\epsilon-16\tau_2
\\
a_1^4a_2^4a_3^4&&32&=32\\
etc.\notag
\end{align}

\end{enumerate}

\end{document}